\documentclass[sigplan,10pt]{acmart}

\renewcommand\footnotetextcopyrightpermission[1]{}
\renewcommand\footnotetextcopyrightpermission[1]{}

\usepackage[english]{babel}
\usepackage{blindtext}
\usepackage{enumitem}
\usepackage{siunitx}
\usepackage[labelfont=normalfont, font=normalfont]{subcaption}
\usepackage{array}
\usepackage[T1]{fontenc}
\def\regioncount{123\xspace}
\def\globalavg{368.39 \carbonunit\xspace}
\def\carbonunit{g$\cdot$CO$_2$eq/kWh\xspace}
\def\emissionunit{g$\cdot$CO$_2$eq\xspace}

\usepackage[many,theorems,skins,breakable]{tcolorbox}
\definecolor{tcbcolback}{RGB}{245,243,253}
\definecolor{tcbcolbox}{RGB}{254 218 173}
\definecolor{tcbcolframe}{RGB}{245,243,253}
\newtcolorbox{visionbox}[2][]{%
    colback=white!12,
    coltitle=black,
    colframe=teal!50,
    fonttitle=\bfseries,
    title=#2, 
    sharp corners,
    rounded corners=southeast,
    boxrule=0pt,
    enhanced,
    drop fuzzy shadow,
    #1, 
    top=3pt,bottom=2pt,left=3pt,right=3pt
    }

\title[Carbon-Aware Spatial and Temporal Workload Shifting]{On the Limitations of Carbon-Aware Temporal and Spatial Workload Shifting in the Cloud}
\titlenote{This work will appear at EuroSys'24.}

\author{Thanathorn Sukprasert}
\affiliation{%
  \institution{University of Massachusetts Amherst}
  \country{USA}
  }

\author{Abel Souza}
\affiliation{%
  \institution{University of Massachusetts Amherst}
  \country{USA}
  }

\author{Noman Bashir}
\affiliation{%
  \institution{Massachusetts Institute of Technology}
  \country{USA}
  }

\author{David Irwin}
\affiliation{%
  \institution{University of Massachusetts Amherst}
  \country{USA}
  }

\author{Prashant Shenoy}
\affiliation{%
  \institution{University of Massachusetts Amherst}
  \country{USA}
  }

\renewcommand{\shortauthors}{Sukprasert et al.}

\begin{abstract}
Cloud platforms have been focusing on reducing their carbon emissions by shifting workloads across time and locations to when and where low-carbon energy is available.
Despite the prominence of this idea, prior work has only quantified the potential of spatiotemporal workload shifting in narrow settings, i.e., for specific workloads in select regions. In particular, there has been limited work on quantifying an upper bound on the ideal and practical benefits of carbon-aware spatiotemporal workload shifting for a wide range of cloud workloads. 
To address the problem, we conduct a detailed data-driven analysis to understand the benefits and limitations of carbon-aware spatiotemporal scheduling for cloud workloads. We utilize carbon intensity data from 123 regions, encompassing most major cloud sites, to analyze two broad classes of workloads---batch and interactive---and their various characteristics, e.g., job duration, deadlines, and SLOs.
Our findings show that while spatiotemporal workload shifting can reduce workloads' carbon emissions, the practical upper bounds of these carbon reductions are currently limited and far from ideal. We also show that simple scheduling policies often yield most of these reductions, with more sophisticated techniques yielding little additional benefit. Notably, we also find that the benefit of carbon-aware workload scheduling relative to carbon-agnostic scheduling will decrease as the energy supply becomes ``greener."
\end{abstract}


\acmConference[EuroSys '24]{Nineteenth European Conference on Computer Systems}{April 22--25, 2024}{Athens, Greece}

\begin{document}

\maketitle
\sloppy

\section{Introduction}
\label{sec:introduction}
The demand for computing continues to increase rapidly and is expected to accelerate further with the mainstream adoption of machine learning (ML) and artificial intelligence (AI) applications, such as ChatGPT~\cite{chat-gpt-paper} and its derivatives. Since computation requires energy, computing's energy consumption is also expected to accelerate in the coming decades. For example, recent estimates project that datacenter energy consumption will increase by at least 10\% per year until 2030~\cite{datacenter-demand}, which is significantly higher than the 1.65\% estimated increase per year in the 2010s~\cite{masanet}. Given these trends, there is an increasing concern that this substantial growth in computing's energy consumption will lead to a proportionate increase in its carbon emissions. Technology companies have recognized this problem and are addressing it by setting aggressive targets for reducing computing's carbon footprint, e.g., such as achieving net-zero emissions by 2030 or even earlier~\cite{amazon-carbon-neutral,facebook-carbon-neutral,vmware-carbon,google-carbon-free,microsoft-carbon-negative, google-blog}.

To achieve the aggressive carbon reduction goals above, researchers have begun to focus on optimizing computing's \emph{carbon-efficiency}, or computations per unit of carbon emitted~\cite{enabling-socc21,ecovisor}, in addition to its energy efficiency, or computations per joule of energy consumed. 
While optimizing for energy efficiency reduces carbon emissions, the benefits will likely be limited moving forward as computing is already highly energy-efficient. Thus, to optimize carbon-efficiency, recent work has focused on leveraging real-time variations in energy's carbon-intensity across time and space by shifting computation to when and where lower-carbon energy is available. The recent emergence of third-party carbon information services~\cite{electricity-map, watttime}, which provide real-time data on energy's carbon-intensity at high temporal and spatial resolution, have enabled this approach. Cloud platforms have used these services to develop tools that provide coarse, high-level per-region estimates of energy’s carbon-intensity~\cite {google-dashboard}. This has led recent work to propose a range of spatial and temporal workload shifting policies that leverage energy's carbon-intensity variations to reduce computing's carbon emissions~\cite{ecovisor,gupta2021chasing,dean-carbon,carbonexplorer,wait-awhile, radovanovic2021carbonaware,hanafy2023carbonscaler,zhou2013carbon}.

\begin{figure}[t]
	\begin{minipage}{0.5\linewidth}
		\includegraphics[width=\linewidth]{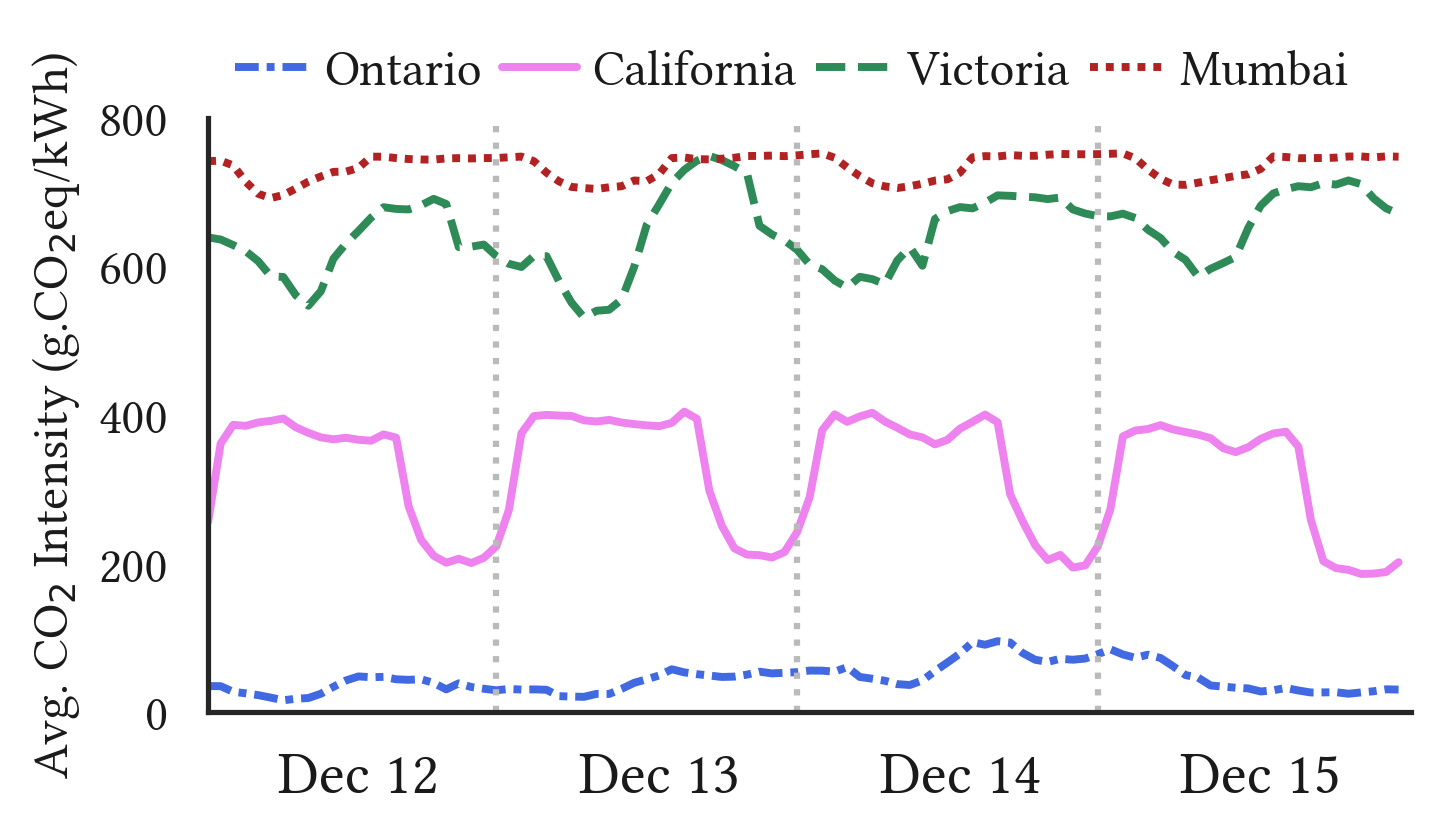}
        \vspace{-0.6cm}
		\subcaption{Carbon Trace}
	\end{minipage}\hfill
	\begin{minipage}{0.5\linewidth}
		\includegraphics[width=\linewidth]{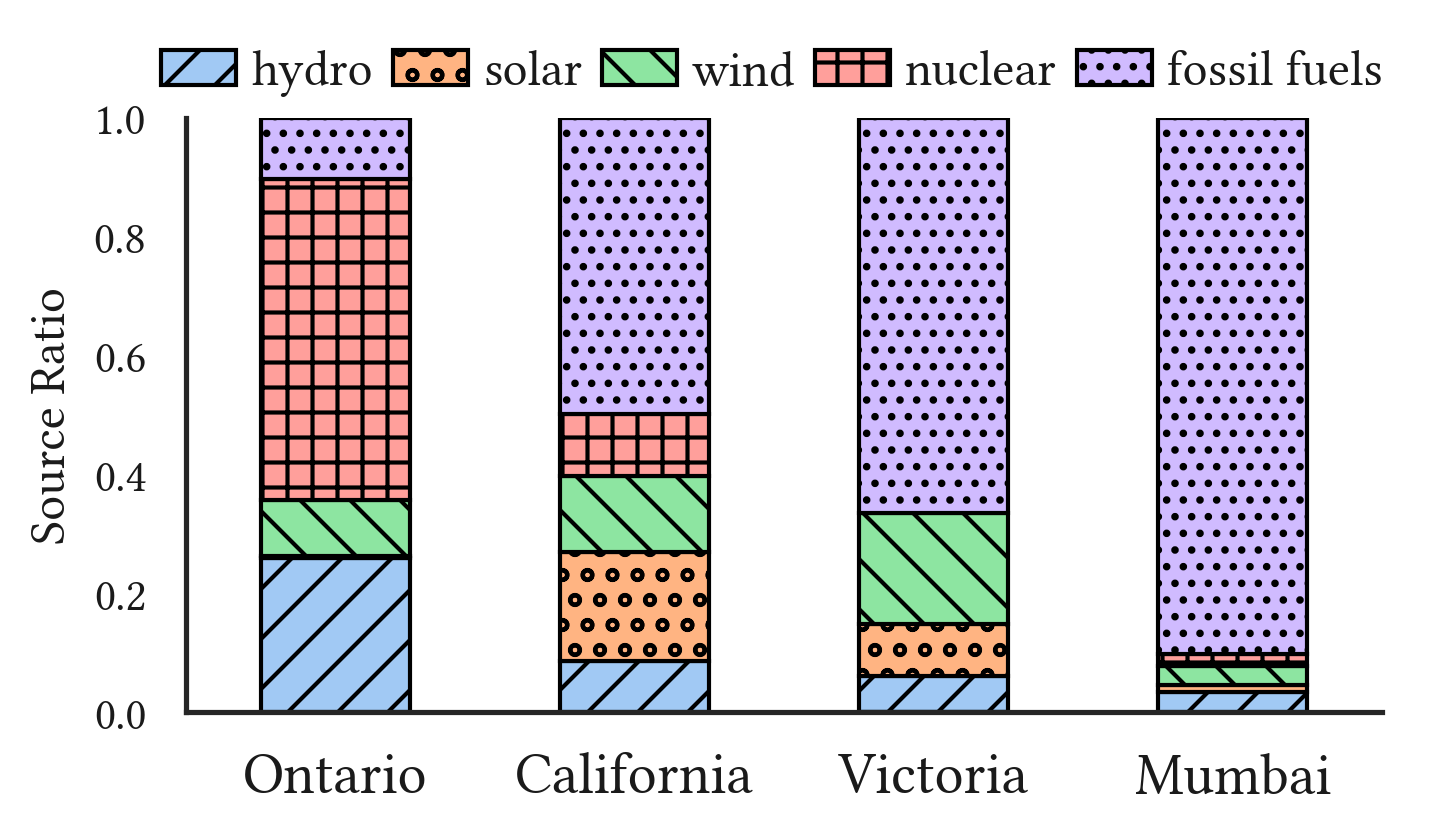}
        \vspace{-0.6cm}
		\subcaption{Generation Mix}
	\end{minipage}
	\vspace{-0.3cm}
	\caption{\emph{The carbon-intensity of energy supplied by the electric grid depends on the grid's energy mix and can vary by 2$\times$ and 43$\times$, temporally and spatially, respectively.}}   
	\label{fig:intro-figures}
\end{figure}

To illustrate, Figure~\ref{fig:intro-figures}(a) shows that grid energy's carbon-intensity can vary by 2$\times$ over a day (in California) and by over 43$\times$ across regions (between Ontario and Mumbai). The magnitude and variability of grid energy's carbon-intensity depend on the mix of energy sources. Traditional fossil fuel-based energy sources, such as coal and natural gas, tend to exhibit high carbon-intensity with low variance. In contrast, renewable sources like solar and wind have low carbon emissions but with highly variable generation. As a result, the proportion of energy generated by renewables versus fossil fuels at any location dictates the magnitude and variance of its energy's carbon-intensity.  Figure~\ref{fig:intro-figures}(b) illustrates this point by showing each region's energy generation mix. California's energy mix is comprised of 50\% renewable energy; hence, the carbon-intensity in California has a low average with high variability. In contrast, $\sim$90\% of Mumbai's energy derives from burning fossil fuels, which results in high average carbon-intensity with low variability. 

Leveraging spatiotemporal variations in energy's carbon-intensity requires workload flexibility. Fortunately, most computing workloads have significant performance, temporal, and spatial flexibility that enable the workloads' intensity, run time, and execution location to be adjusted according to the availability of low-carbon energy~\cite{jahanshahi2022powermorph}. For instance, batch machine learning (ML) training jobs often have substantial temporal flexibility that enables them to be \emph{suspended} during high-carbon periods and \emph{resumed} during low-carbon periods~\cite{wait-awhile}. Similarly, interactive inference requests for object detection may have spatial flexibility that enables migrating them to a location with low carbon-intensity~\cite{igsc2023casper}. These insights have led to an implicit assumption within the systems research community that schedulers can harness such workload flexibility to significantly reduce carbon emissions~\cite{jahanshahi2022powermorph}.

While workload flexibility enables computer systems to exploit low-carbon energy, the potential for reducing carbon from carbon-aware spatiotemporal workload shifting ultimately depends on the variability in the carbon-intensity of energy across time and regions. In particular, temporally shifting workloads is less effective when a location, such as Mumbai, has few variations in carbon-intensity. Similarly, spatial shifting is less effective if the rank order of regions by carbon-intensity never changes, as shown in Figure~\ref{fig:intro-figures}(b). In this case, the optimal spatial shifting policy is simply to run the job in the lowest carbon region. Of course, external constraints beyond carbon-intensity, such as capacity constraints, latency requirements and privacy regulations, such as GDPR~\cite{gdpr}, may also affect spatial shifting decisions. 

Understanding and quantifying the potential carbon reductions from spatiotemporal workload shifting is crucial for informing ongoing research efforts. While there has been preliminary work on this topic~\cite{mccallum,emma2,wait-awhile}, most analyses focus on a specific setting, e.g., batch ML training jobs, using a small number of geographical regions. Thus, the potential for reducing carbon with such approaches is not clear. 

To address the problem, we conduct a large-scale analysis of worldwide carbon-intensity data, while varying cloud workload characteristics, to quantify an upper bound on carbon reductions from spatiotemporal workload shifting.
Our analyses use a carbon trace dataset that includes \regioncount regions worldwide, where each region's data includes hourly carbon-intensity over three years (2020--2022). We then quantify an upper bound on carbon reductions under ideal, unconstrained conditions, as well as analyze how simple practical constraints on workload flexibility affect these carbon reductions.
\emph{Our primary finding is that while computer systems can reduce their carbon emissions using spatiotemporal workload shifting, the ideal carbon reductions are limited, and practical constraints in workload flexibility significantly reduce the ideal savings}. Thus, based on our analysis and contrary to conventional wisdom, we conclude that i) carbon-aware spatiotemporal workload shifting is likely not a panacea for significantly reducing cloud platforms' carbon emissions, and ii) sophisticated policies, in many cases, yield little additional benefits as compared to simple policies.
Below, we distill the key points of our analysis.

\begin{itemize}[leftmargin=0.35cm, itemsep=0.07cm, topsep=0.05cm]

\item Currently, more than 70\% of regions worldwide have low daily carbon-intensity variations\footnote{Low daily variations refer to a coefficient of variation less than 0.1.} due to their reliance on fossil fuel-based, hydro, or nuclear power generation. These energy sources have low variability, which limits the benefit of temporal shifting in many cloud regions.

\item While the ideal potential for reducing carbon emissions via spatial migration is significant (up to 352 \emissionunit), the practical potential is likely much less ($<$115 \emissionunit) due to practical constraints, such as latency requirements, capacity limitations, and privacy regulations. 

\item The potential carbon reductions derived from real-world cloud workloads are limited (<112 \emissionunit). 
The primary reason is that large jobs, which tend to have the least temporal flexibility, are a substantial fraction of cloud data centers' computing workloads.

\item Currently, regions' carbon-intensity maintains the same rank order most of the time. Thus, there is little need for sophisticated migration policies, as migrating to the lowest-carbon region once maximizes carbon reductions.

\item While the ideal potential for reducing carbon from temporal shifting can be as much as 189 g$\cdot$CO$_2$eq
in some cases, practical constraints limit this to 32 g$\cdot$CO$_2$eq on average. 

\item Carbon reductions from spatial shifting are substantially higher than those from temporal shifting. 
\item Finally, the benefits of carbon-aware workload scheduling relative to carbon-agnostic scheduling will decrease as the world's energy supply becomes ``greener."
\end{itemize}

\vspace{0.05cm}
We present a comprehensive analysis of carbon reductions from spatiotemporal workload shifting. In particular, we shed light on critical takeaways for designing carbon-aware policies for scheduling workloads. Our insights have important implications for the design of carbon-aware systems, an area of increasing importance in cloud computing and in today's environmentally-conscious landscape.

\section{Background}
\label{sec:background}
Below, we provide an overview of our carbon-intensity data, as well as the workload characteristics that affect carbon-aware spatiotemporal workload shifting.  
\vspace{-0.2cm}

\subsection{Grid Energy's Carbon-intensity}

Our analysis uses historical time-series data of grid energy's carbon-intensity (in \carbonunit) from \regioncount regions worldwide. As mentioned in \S\ref{sec:introduction}, grid energy's carbon-intensity changes over time based on the mix of generators required to satisfy a variable demand. The electric grid's energy demand varies based on human behavioral patterns, e.g., day/night, weekday/weekend, etc., and weather, which influences the energy necessary for indoor heating and cooling. The grid is divided into different regions operated by their own balancing authority, called Independent System Operators (ISOs) and Regional Transmission Organizations (RTOs), which must satisfy a region's demand using a variety of generators with different characteristics, e.g., fuel types, capacities, ramp rates, and, importantly, carbon-intensities. 

Energy's average carbon-intensity for a specific region at any moment is the average carbon-intensity for each of its generators weighted by their energy generation. Many generation sources, including nuclear, geothermal, hydroelectric, solar, and wind, have low carbon-intensity. In addition, solar and wind energy, which are the fastest growing energy sources~\cite{usatoday}, are ``non-dispatchable,'' i.e., their generation is uncontrollable. In contrast, fossil fuel-based generation, such as coal, oil, and natural gas, have higher carbon-intensity. Each region's mix of generators also differs based on its unique climate and access to natural resources. For example, while some regions have abundant hydropower due to the presence of large rivers, such as in the northwest U.S., others have abundant solar power, such as in the southwest U.S. In addition, the variability in grid-tied solar and wind's low-carbon energy output manifests as variations in grid energy's carbon-intensity. Thus, regions with more solar and wind tend to have more carbon-intensity variations. 

Until recently, the carbon-intensity of grid energy was opaque to consumers since energy generation data was not easily available. However, balancing authorities have begun publicly releasing information about the active generator set and their real-time energy output via web APIs. Carbon information services, such as Electricity Map~\cite{electricity-map} and WattTime~\cite{watttime}, combine the grid's real-time generation information with models based on each generator's characteristics to infer grid energy's real-time carbon-intensity in each region and make it available via web APIs. As discussed in \S\ref{sec:methodology}, we collect multiple years of this data for our analysis.  

Note that our analysis focuses narrowly on grid energy's \emph{average carbon-intensity}, which falls under Scope 2 emissions in the GHG protocol\footnote{Scope 1 emissions occur when an organization directly burns fossil fuels, e.g., in backup diesel generators, and other chemicals.}~\cite{ghg-protocol}. The use of grid energy accounts for the vast majority of datacenters' operational emissions, which include Scopes 1 and 2. Importantly, we do not analyze Scope 3 emissions for either datacenters or the electric grid. Scope 3 mostly covers embodied carbon emissions that result from the production of the products and services companies use.  For example, building a datacenter or power plant also incurs carbon emissions. While Scope 3 emissions are important, they are more challenging to measure and optimize accurately, e.g., by increasing server lifetime, selecting ``greener'' suppliers, etc., and generally have a less direct effect on operations. In addition, we focus our analysis on average carbon emissions rather than marginal carbon emissions since the GHG protocol only requires reporting the former.  Energy's marginal carbon-intensity is the carbon-intensity of satisfying the next unit of energy demand. The GHG protocol does not use marginal carbon-intensity primarily because accurately measuring it is difficult and requires a precise knowledge of when each generator and load starts and stops. 
\subsection{Spatiotemporal Workload Flexibility}
Cloud datacenters serve two broad classes of workloads -- batch and interactive -- each with its own dimensions and degrees of spatiotemporal flexibility.

\subsubsection{\textbf{Batch Workloads.}}
Common batch workloads include data processing tasks, machine learning training, scientific computing, and simulations. Such workloads desire high throughput and often do not have strict latency requirements. Consequently, batch workloads generally include jobs with some degree of ``slack'' and thus may be delayed, or interrupted, although not indefinitely. Schedulers often exploit this slack by deferring batch jobs' start time, i.e., forcing them to wait in a queue or periodically interrupting their execution. Cluster schedulers, such as Google's Borg~\cite{burns2016borg}, Kubernetes~\cite{burns2016borg}, and Slurm~\cite{slurm02}, often defer or interrupt batch jobs to satisfy higher-priority requests, either by terminating low priority jobs or checkpointing their state and resuming them later. Such schedulers can also defer or interrupt jobs when energy's carbon-intensity is high to lower carbon emissions. In addition to temporally shifting batch jobs, they can be spatially migrated to regions with lower carbon-intensity. While deferring, interrupting, and spatially migrating jobs can incur overhead, which depends on the size of a batch workload's memory and disk state, we present optimistic analyses that ignore such overheads to provide an upper bound on carbon reductions from migrating batch jobs. As mentioned in \S\ref{sec:introduction}, we do consider how regulatory policies, such as HIPPA and GDPR, may hinder jobs' spatial migration outside their local region. 

\subsubsection{\textbf{Interactive Workloads.}}
Interactive workloads include small server requests, such as web and inference requests, that require a low-latency response. While such requests cannot be delayed and have no temporal flexibility, they can often be flexibly routed to different datacenters for servicing. For example, prior work has proposed policies for migrating (or routing) web requests to datacenters in global Content Distribution Networks (CDNs) based on datacenter load, electricity prices~\cite{cdn2}, and energy~\cite{cdn3,cdn1}. Nevertheless, the ability to redirect requests may be restricted, and in certain cases, it may not be feasible for interactive workloads. This is particularly true considering regulatory~\cite{gdpr} and latency~\cite{igsc2023casper} constraints, as well as the extensive preparations needed to facilitate accurate traffic re-routing, including state replication across multiple datacenters~\cite{yang2023skypilot}.

\section{Objectives and Methodology}
\label{sec:methodology}
The primary goal of our analysis is to quantify an upper bound on carbon reduction from spatiotemporal workload shifting under ideal and constrained conditions. Our hypothesis is that while the upper bound of spatiotemporal workload shifting exhibits significant reductions in computing’s carbon emissions, a substantial gap exists between the ideal and constrained conditions. To quantify carbon reduction and evaluate our hypothesis, we focus on answering the specific research questions below. We then outline our methodology for answering these questions.

\begin{enumerate}[leftmargin=*, topsep=0.2cm, itemsep=0.07cm]
\item \textbf{Global Carbon Analysis.} What are the characteristics of grid energy's carbon-intensity worldwide? How do its magnitude, variance, and periodicity vary across regions? How has it changed in recent years? (\S\ref{sec:analysis}).

\item \textbf{Spatial Migration:} How much carbon reduction is possible from spatially migrating workloads? How might capacity, latency SLOs, and regional privacy constraints impact this carbon reduction What is the optimal policy for minimizing carbon emissions? (\S\ref{sec:spatial}).

\item \textbf{Temporal Shifting.} How much carbon reduction is possible from temporally shifting delay-tolerant batch workloads? How does this carbon reduction vary with workload characteristics, such as job length and slack? (\S\ref{sec:temporal}).

\item \textbf{What-If Scenarios.} What are the benefits of combining spatial and temporal shifting, and how much carbon reduction accrues from each method? How does the i)
ratio of migratable workload, ii) prediction error, and iii) increase in renewables impact the carbon reductions from temporal and spatial shifting? 
(\S\ref{sec:what-if})
\end{enumerate}

\noindent As discussed in later sections, our real-world trace-driven analysis shows that \emph{while the upper bound on carbon reduction can be substantial, there exists a significant difference between the ideal and constrained settings.} This leads to additional questions below on the implications for systems design.

\begin{enumerate}[leftmargin=*]
\setcounter{enumi}{4}
\item\textbf{Implications:} What are the implications of our analysis for cloud operators and the systems research community? 
How can our analysis guide carbon optimizations in both current and future grids? (\S\ref{sec:analysis}-\S\ref{sec:what-if}).
\end{enumerate}\

\subsection{Analysis Setup}

Below, we provide details on our i) carbon-intensity data sources, ii) workload characteristics, iii) approach to derive carbon emissions with and without leveraging workload flexibility, and iv) metric for quantifying carbon reduction. 

\subsubsection{\textbf{Carbon-intensity Data}}
We collected carbon-intensity traces for \regioncount different geographical regions worldwide from 2020 to 2022 using the Electricity Maps~\cite{electricity-map} web API. Each trace reports energy's average carbon-intensity, measured in grams of carbon dioxide equivalent per kilowatt-hour (\carbonunit), in hourly granularity. The hourly granularity is the highest granularity for average carbon-intensity data currently available from public sources. As shown in \S\ref{sec:analysis}, since grid energy's carbon-intensity rarely varies significantly within a 2-3 hour period, higher granularity data would likely not change the results of our analysis. The \regioncount locations include our entire carbon trace dataset and encompasse 99 known datacenter locations: 35 for Google Cloud Platform (GCP), 24 for Microsoft Azure, 23 for Amazon Web Services (AWS), 7 for IBM, and 10 for Alibaba.  

\begin{table}[t]
\footnotesize
\begin{center}
\begin{tabular}{|| l | c | c ||} \hline
{\bf Dimension} & {\bf Range / Description} & {\bf Source} \\ \hline \hline 
{\emph{Type}} & Batch, interactive & \cite{borg-next-gen}   \\ \hline
{\emph{Length (Hour)}} & 0.01, 1, 6, 12, 24, 48, 96, 168 & \cite{borg-next-gen}  \\ \hline
{\emph{Deferrability}} & 24H, 7D, 24D, 30D, 1Y, 10$\times$& \\ \hline
{\emph{Interruptibility}} & Zero overhead & \\ \hline
{\emph{Spatial Migration}} & Zero overhead & \\ \hline
{\emph{Job Arrival Time}} & Every hour of the year &  \\ \hline
{\emph{Job Origin}} & \regioncount locations & \cite{electricity-map} \\ \hline 
{\emph{Resource Usage}} & Energy-optimized 100\% usage &  \cite{borg-next-gen, azure}  \\ \hline
\hline  
\end{tabular}
\vspace{0.1cm}
\caption{\emph{Workload characteristics, flexibility dimensions and degrees, and configurations.}}
\label{tab:workload-character}
\end{center}
\vspace{-0.7cm}
\end{table}

\subsubsection{\textbf{Workload Configuration}}

Table~\ref{tab:workload-character} outlines the workload configurations for our analysis. We define workload characteristics across these dimensions as the following.

\begin{enumerate}[leftmargin=*, topsep=0.08cm, itemsep=0.1cm]
\item \textbf{Job Length} is the amount of time a job needs to complete its execution without interruption. We examine a range of job lengths that map to interactive jobs (1 minute or less), small batch jobs (1hr to 24hrs), long batch jobs (24-168hrs), and uninterruptible service jobs ($>$168hrs). The range of job lengths and values within that range are based on version 3 of Google's Borg cluster trace~\cite{borg-next-gen,bashir2021take}.

\item \textbf{Job arrival time} is the submission time of a job. Since jobs can arrive at any point of the day, we consider all the possible job starting times. As our trace is collected at hourly granularity, jobs can start at the hour boundary, and there are 8760 potential start times over a year. We compute carbon reductions at all start times and report the average results and confidence intervals. 

\item \textbf{Spatial Migration} defines the spatial flexibility of a job. When a job is migrated from one region to another, it can incur overheads depending on the size of its memory and disk state. Our analysis ignores these migration overheads in quantifying an upper bound on carbon reduction.

\item \textbf{Deferrability} characterizes the temporal flexibility of a job to delay the start of its  execution based on its \emph{slack}, which dictates the maximum delay possible. Prior work suggests that practical waiting times for batch jobs are generally between a few minutes to less than 24 hours~\cite{pradeep:sc20,rodrigo2018towards}. However, to quantify an upper bound on carbon reductions, we vary our slack from 24 hours to a year in quantifying an ideal upper bound on carbon reduction.


\item \textbf{Interruptibility} defines the performance flexibility of a job. An interruptible job can be paused and resumed without a significant loss of computation. While suspending and resuming a job incurs some time and energy overhead (based on an application's memory footprint) that increases emissions, our analysis ignores it since we focus on quantifying an upper bound on carbon reduction. 

\end{enumerate}

\begin{figure}[t]
    \begin{minipage}{0.5\linewidth}
        \includegraphics[width=\linewidth]{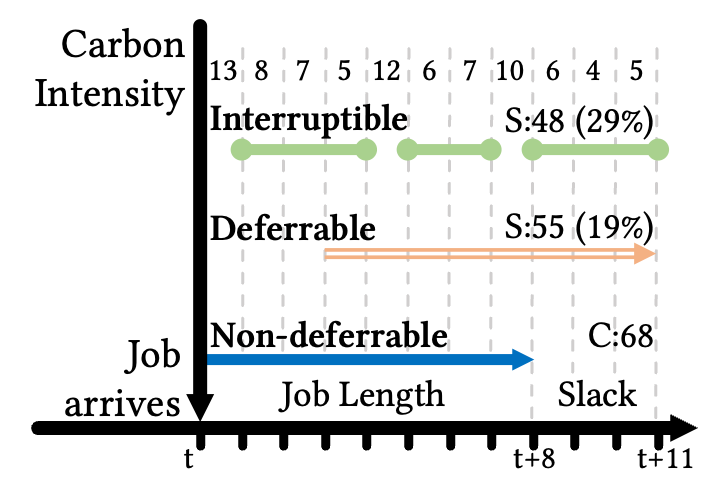}
        \subcaption{Temporal Shifting}
    \end{minipage}\hfill
    \begin{minipage}{0.5\linewidth}
        \includegraphics[width=\linewidth]{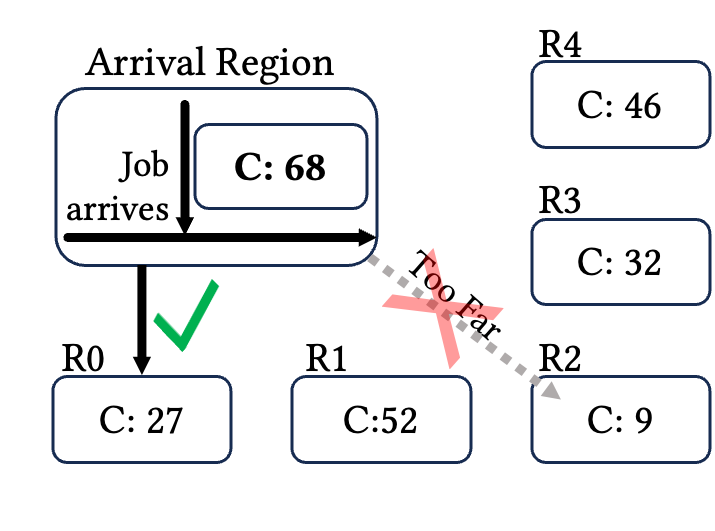}
        \subcaption{Spatial Migration}
    \end{minipage}
    \caption{\emph{Temporal shifting and spatial migration illustration.}}   
    \vspace{-0.8cm}
    \label{fig:temporal-spatial-illustration}
\end{figure}


\subsubsection{\textbf{Metrics.}} We quantify carbon reduction in terms of \emph{absolute carbon reduction}, which captures how many grams of carbon are reduced from spatiotemporal workload shifting. We also present carbon reductions as \emph{global average carbon reduction} to quantify how much the absolute carbon reduction is relative to the world's average carbon intensity. Below, we define how both metrics are calculated. 

\begin{enumerate}[leftmargin=*, itemsep=0.1cm, topsep=0.08cm]
\item[a)] \textbf{Absolute Carbon Reduction} is the difference between carbon emissions after any spatiotemporal workload shifting and the carbon-agnostic baseline. We measure it in \emissionunit, where a higher value is better. 

\item[b)] \textbf{Global Average Reduction} is the amount of average absolute carbon reduction of \regioncount regions from spatiotemporal workload shifting as a percentage of the global average carbon-intensity of \globalavg.  
\end{enumerate}

\subsection{Analysis Workflow}
Below, we describe the high-level workflow for each of our temporal and spatial workload shifting policies.

\begin{figure*}[t]
    \centering
    \begin{tabular}{cc}
    \includegraphics[width=0.47\linewidth]{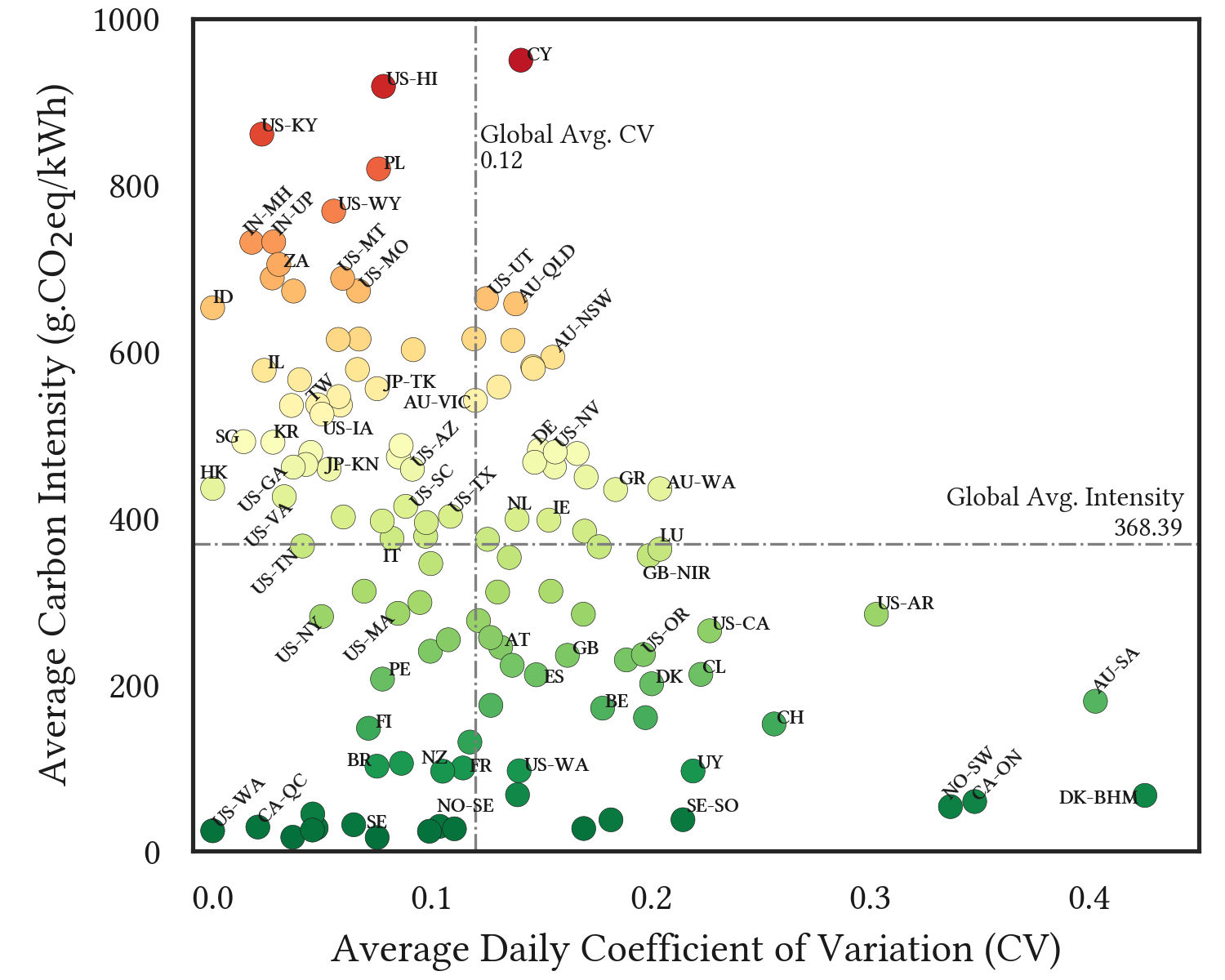} &
    \includegraphics[width=0.47\linewidth]{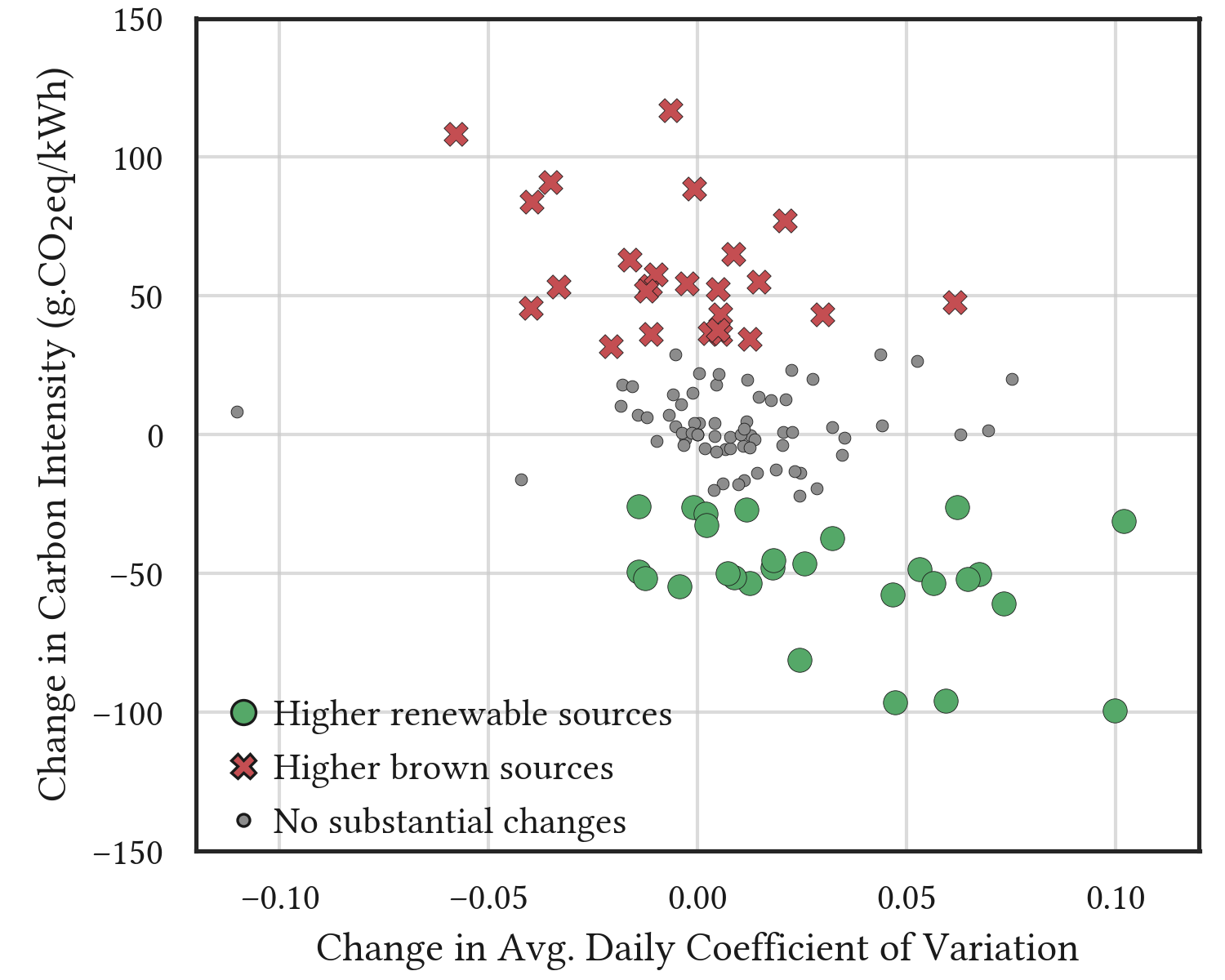} \\
    (a) Mean and CV & (b) Change Over Time 
    \end{tabular}
    \vspace{-0.4cm}
    \caption{\emph{Average carbon-intensity and average daily variability in 2022 as well as change in average carbon-intensity and daily variability from 2020 to 2022. A negative change indicates the value of the variable decreased over time.}}   
    \label{fig:analysis1}
\end{figure*}
\subsubsection{\textbf{Temporal Workload Shifting.}}

In our temporal analysis, the different dimensions of a job include its length, slack, and the ability to defer and interrupt it. Here, we assume perfect knowledge of the future carbon-intensity and job length. Figure \ref{fig:temporal-spatial-illustration}(a) shows our methodology for computing carbon reduction under different dimensions. In a non-deferrable or baseline scenario, a job arrives at time $t$ and immediately starts running. In our toy example, such an execution yields 68 units of carbon emissions. If the job is deferrable, we find contiguous slots with minimum cumulative carbon emissions that can fit a given job. Our problem maps well to the standard k-element sub-array with minimum sum~\cite{k-subarray}, where the length of the array is equal to the sum of job length and slack. In our example, the job is deferred to provide 13 units of absolute reduction and 19\% relative reduction. For interruptible jobs, we find k minimum elements in an array, considering no overhead, to find the slots that can finish the job. In the example, the interruptibility leads to absolute reduction of 48 units and relative reduction of 29\%.  We repeat this analysis for all the job arrival times and present the mean and standard deviation.


\subsubsection{\textbf{Spatial Workload Shifting.}} 

In our spatial analysis, the different dimensions are the job length, regions the job can migrate to, and the policy it uses to decide on the migration. As with our temporal analysis, we assume we know the job length and carbon-intensity across all the regions in the world. Figure~\ref{fig:temporal-spatial-illustration}(b) shows how spatial shifting works. A job arrives in a given region, and without any spatial migration, its carbon emissions are 68 units. However, during the same start time, the job would have a different amount of carbon emissions if it migrates to one of the other regions. Under a ``global'' setting, the job can migrate to any region in the world, including the R2 region in the figure. However, a job may have a latency target or regulations that prevent it from moving to the greenest region. As a result, it may only be able to migrate to R0, which reduces its reduction from spatial shifting. We define various regions in our analysis based on the geographical boundaries. The carbon reduction is calculated similar to that of the temporal analysis. 

\section{Global Carbon Analysis}
\label{sec:analysis}

To understand the limits of reducing carbon from spatiotemporal workload shifting, we first analyze the carbon-intensity signal of regions worldwide to recognize their magnitude, variation, patterns, and changes over time. 

\subsection{Carbon's Magnitude and Variance}
\label{subsec:daily_mean_cv}

Since spatiotemporal workload shifting exploits differences in grids' carbon-intensity across time and regions, we analyze carbon traces from \regioncount regions across all five continents to understand their characteristics. Note that many of these regions have cloud datacenters from hyperscale cloud providers. Our analysis examines the average carbon-intensity of each region and hourly variations in the carbon-intensity, expressed as the coefficient of variation\footnote{Variability is quantified using the coefficient of variation, which is computed as the standard deviation divided by the mean.} of each region's carbon-intensity. Figure \ref{fig:analysis1}(a) depicts the average and the coefficient of variation (CV) of each region. The dotted lines in the figure represent the average carbon-intensity (CI) and the average CV across all regions. As shown, the dotted lines also partition the figure into 4 quadrants, that represent four combinations of low and high CI and low and high CV, e.g., the bottom left quadrant is the low-low quadrant where regions have low CI and low CV and so on.

Overall, electric grids that largely depend on fossil fuels to generate electricity have above average (``high'') carbon-intensity, while regions with a high degree of renewable sources such as hydro, geothermal, solar, and wind have below average (``low'') carbon-intensity. Similarly, since electricity generation from fossil fuels tends to be stable over time, grids that derive electricity from such sources exhibit low variations in carbon-intensity. Conversely, grids with higher fractions of intermittent sources, such as wind and solar, exhibit higher variations in hourly carbon-intensity. 

Figure \ref{fig:analysis1} reveals several insights. In particular, the data shows that a significant number (46\%) of regions have above-average carbon-intensity ($> 400$ g$\cdot$CO$_2$e/kWh) due to a high reliance on brown sources in many parts of the world. At the same time, many (54\%) of the regions also have below average ($< 400$ g$\cdot$CO$_2$e/kWh) carbon-intensity. This implies that workloads from regions that lie in the top two quadrants will benefit from spatial workload shifting by migrating workloads to cloud regions in the bottom two quadrants, which can potentially yield significant emission reductions. For example, there is a 40$\times$ difference in average CI between the highest and lowest regions. The figure also reveals that a majority of regions have below average CV (i.e., lie in the left half), with fewer than 43\% of the regions with above average CV. As we will see, the lower the CV, the less effective temporal workload shifting (\S\ref{sec:temporal}). Many regions with a low carbon-intensity also have a low CV due to stable, clean generation, e.g., from nuclear, hydro, and geothermal. Even for many regions in the bottom two quadrants (with below-average CI due to using clean energy), relatively few lie in the right half of the figure, with above-average CV. 

\begin{figure*}[t]
	\centering
	\includegraphics[width=\linewidth]{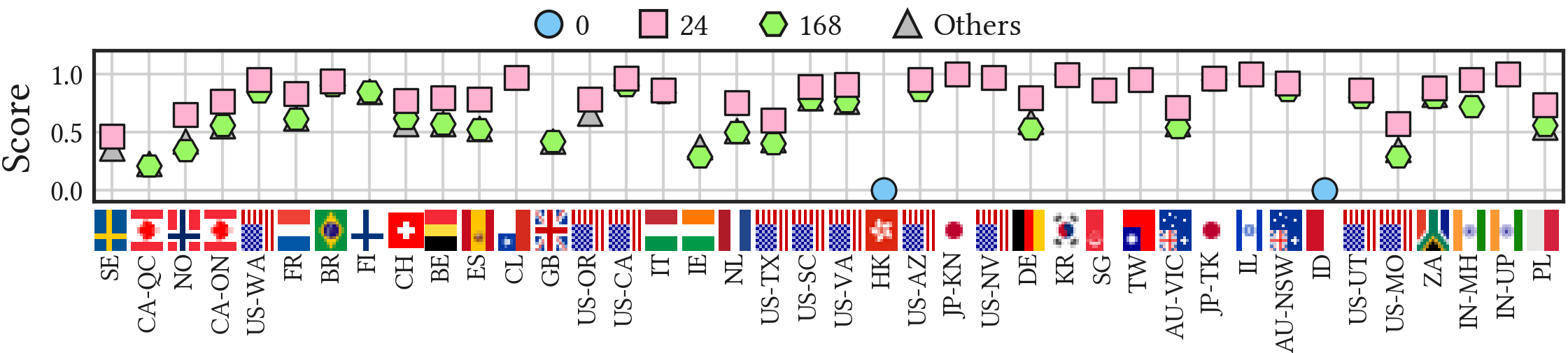}
	\vspace{-0.8cm}
	\caption[]{\emph{Periodicity score for 40 regions with datacenters (AWS, Azure, GCP), the two most common periods are 24- and 168-hour periods. The regions are ordered by their average carbon-intensity from lowest to highest.}}
	\label{fig:periodicity_score}
\end{figure*}

Thus few regions will see significant benefits from temporal shifting methods. Of the regions in the right half, those in the high-high quadrants will see the highest absolute reductions. Overall, the clustering of regions in various quadrants indicates that global carbon-intensity has a medium average magnitude but varies widely, with a low average variance.

\subsection{Long Term Trends}
\label{subsec:longterm_trends}
The carbon-intensity of the electric grid in each region depends on that region's energy mix and production levels. The mix has changed over time, particularly as various grids attempt to decarbonize and transition to lower carbon sources, e.g., by deploying more solar and wind farms. To understand how grids are evolving in each region, we analyze the changes in energy's carbon-intensity over three years period. Figure~\ref{fig:analysis1}(b) shows the change in each region's average carbon-intensity and CV from 2020 to 2022. Ideally, we would want the change in CI to be negative and the change in CV to be positive, as this would indicate lower emissions due to increased adoption of renewable energy.  

We derive clusters for Figure~\ref{fig:analysis1}(b) using the K-Means++~\cite{arthur2007k} heuristic with input $k=3$ for three different groupings with positive, negative, and no changes in the regions' mix of energy sources. The figure shows that approximately 23\% of regions lowered their carbon-intensity, while the CI actually increased in 20\% of the regions. The increase in CI is because there is an increase in fossil fuel production and decrease in renewable production in these regions. In particular, regions with higher renewable sources generally experience an increase in CV, since most of the renewable sources, besides hydro, have high variability. We consider regions with $\pm$25 g$\cdot$CO$_2$e/kWh to have insignificant changes, and $\sim$57\% of regions fall into this category. Thus, for most regions, there has not been a meaningful change in CI over a 3-year period, indicating that significant changes in grid CI is a slow process. Thus, the conclusions from our analysis are likely to hold for the next several years.  We discuss the implications of increasing renewable production globally in \S\ref{sec:what-if}.


\subsection{Periodicity Analysis}

\label{subsec:periodicity_score}

Apart from the hourly variations, the carbon-intensity of a region also exhibits periodicity at longer time scales. Intuitively, the electricity demand of a grid follows a diurnal pattern across days and nights. Since generation must match demand, the mix of sources used to meet that time-varying demand should follow similar daily patterns, influencing the resulting carbon-intensity. We conducted a time series analysis to estimate the degree of periodicity present in the carbon-intensity trace of each region. The degree of periodicity is expressed as a periodicity score between 0 and 1, with 0 indicating no period exists and 1 indicating that the time series has exactly the same pattern for that particular period. We calculate the periodicity score using Azure Data Explorer's function \texttt{series\_period\_detect()}~\cite{periods-detect}, which uses a Fast Fourier Transform to detect all periods present in a time-series~\cite{periods-detect-fft} and assigns them a score between 0 and 1. 

Figure~\ref{fig:periodicity_score} depicts the detected period within each trace for 40 geographic regions with hyperscale datacenters and their periodicity score. As shown, 35 of the 40 regions (87\%) exhibit a 24-hour period in the carbon-intensity variation with a periodicity score of 0.5 or higher. 
For example, a 24-hour timeframe in US-WA (Washington) has a periodicity score of 1, indicating that the trace's exact pattern repeats daily, often dictated by human-behavior diurnal patterns.
Most of these regions also exhibit a weekly cycle (168-hour period), explained by repeating weekday-weekend effects. 
In contrast, only two regions, namely Hong Kong (HK) and Indonesia (ID), exhibit no periodicity, as indicated by a score of 0. The absence of periodicity, coupled with their high carbon-intensity profiles (Figure \ref{fig:analysis1}), suggests a significant reliance on fossil-fuel sources and the resulting carbon-intensity remaining unchanged over time.
Overall, the presence of daily periods indicates that carbon-intensity values exhibit similar diurnal patterns from one day to the next and are predictable. This periodicity is advantageous for temporal workload shifting because it offers a level of predictability, enabling the scheduling of workloads to low-carbon periods.

\section{Spatial and Temporal Shifting}
\label{sec:spatial-and-temporal}
In this section, we analyze the potential for carbon reduction from spatial and temporal shifting policies, and discuss the implications for systems design.   
\begin{figure*}[t]
	\centering
	\begin{tabular}{ccc}
		\includegraphics[width=0.3\linewidth]{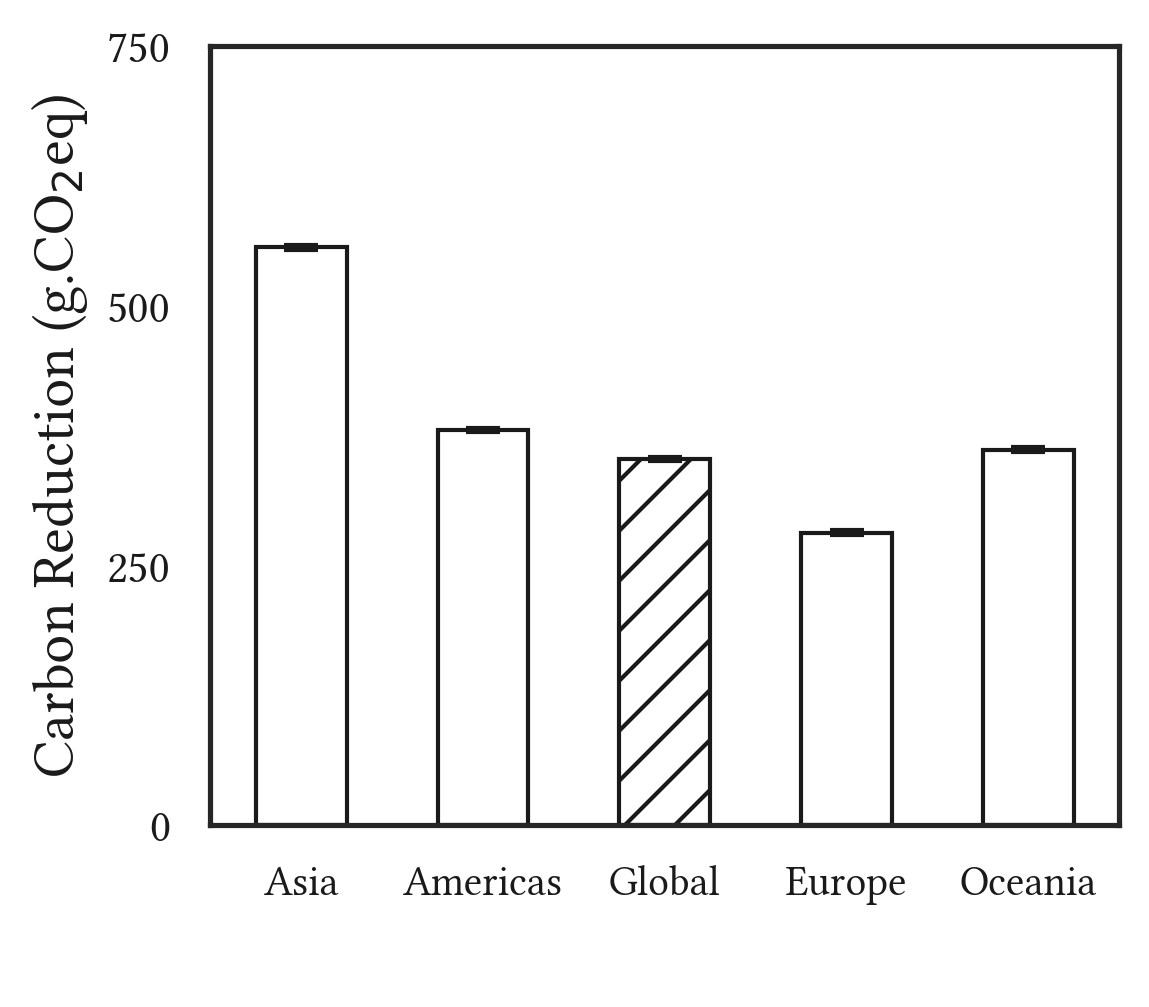} &
		\includegraphics[width=0.3\linewidth]{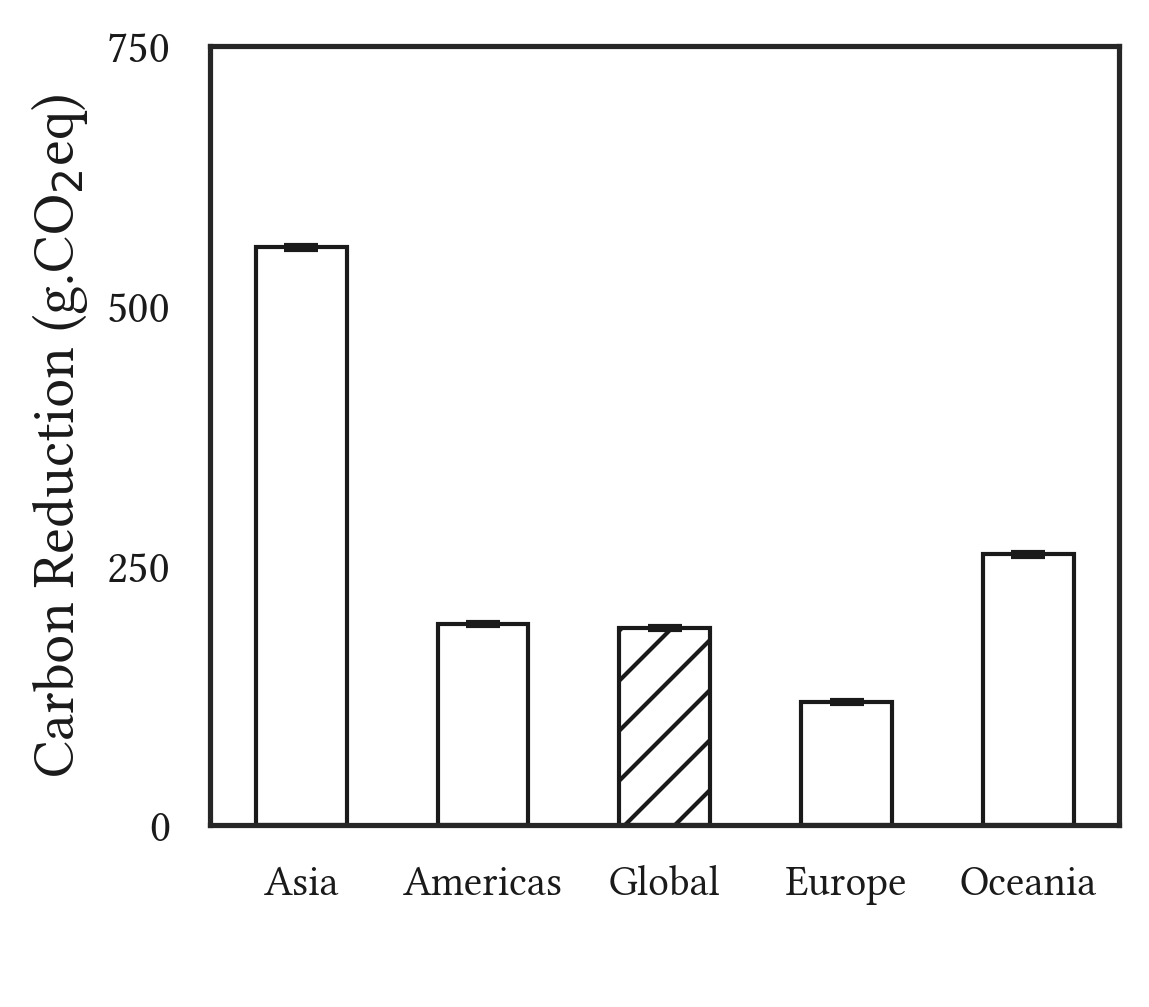} &
		\includegraphics[width=0.3\linewidth]{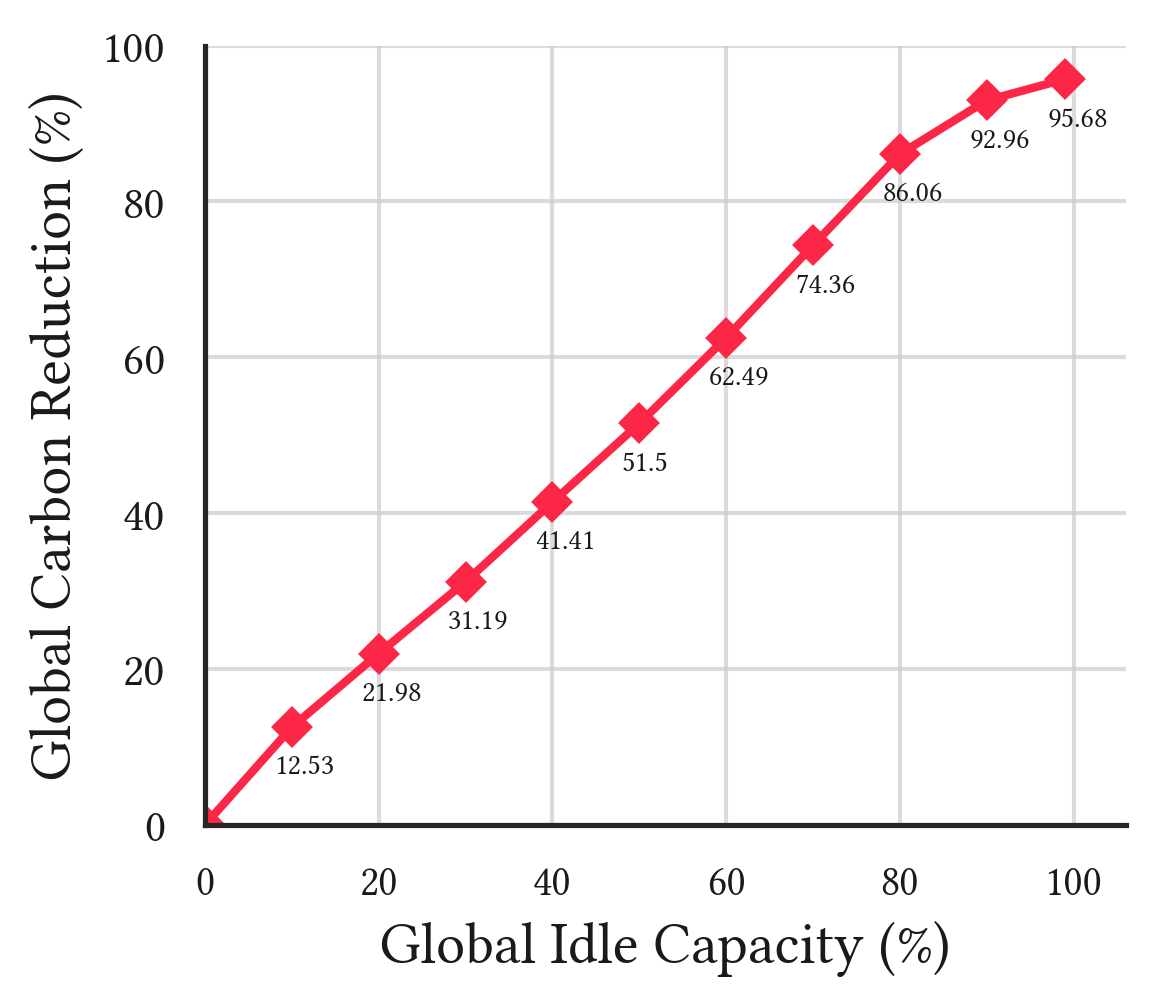}  \\
		(a) Infinite Capacity & (b) With 50\% Utilization per region & (c) Carbon reduction and idle capacity
	\end{tabular}
	\vspace{-0.3cm}
	\caption{\emph{{Carbon reductions with respect to regions' idle capacity for different geographical groupings and with respect to the global average. As the idle capacity increases, the overall carbon reduction increases.}}} 
    \vspace{-0.3cm}
	\label{fig:spatial-cap-figs}
\end{figure*}

\subsection{Spatial Shifting}
\label{sec:spatial}
 We first quantify the upper limits of spatial workload migration by considering an ideal scenario in which all regions have datacenters with infinite capacity and workloads are allowed to migrate anywhere in the world. 
We then analyze how simple constraints such as datacenter capacity, latency, and constrained geographical groupings affect carbon reductions in spatial shifting. 
Last, we discuss sophisticated policies required to optimize for carbon reductions. 

\subsubsection{Carbon Reduction with Infinite Capacity}
\label{subsec:spatial-inf-cap}

In this setting, regions have infinite capacity and workloads can migrate to the greenest region to maximize carbon reductions. In our trace, Sweden has the lowest annual average carbon-intensity in the world at $\sim$16 \carbonunit.  
 
Figure~\ref{fig:spatial-cap-figs}(a) shows the carbon reductions for various geographical groupings. Since all jobs migrate to Sweden, the global average carbon-intensity  drops by 352 g$\cdot$CO$_2$eq, a 96\% reduction when compared to the global average emissions. 
Since the average carbon-intensity varies across different geographical regions, depending on their their respective energy mix, the the extent of carbon reduction from spatial migration also varies significantly. For example, Europe is the greenest overall region in the world with an average carbon-intensity of $\sim$280 \carbonunit. Thus, {its average reduction amounts to only 281 g$\cdot$CO$_2$eq, representing a mere 24\% reduction compared to the global average, which is significantly lower than the global grouping. In contrast, 
Asia has the highest average carbon-intensity of $\sim$540\carbonunit  and as such, its regions experience substantial average reductions of 556 g$\cdot$CO$_2$eq from moving to Sweden, a 115\% reduction relative to the global average emissions.

\vspace{-0.2cm}
\begin{visionbox}{}
\noindent\emph{\textbf{Key Takeaway.} In the ideal case with infinite datacenter capacity, spatial shifting to the world's greenest cloud regions can lead to an average carbon reduction of 352 g$\cdot$CO$_2$eq, i.e., a 96\% reduction relative to the global average carbon emissions. In particular, high intensity regions experience the most benefits when migrating to the greenest locations.}

\end{visionbox}

\subsubsection{Carbon Reduction with Capacity Constraints}
\label{subsec:spatial-cap-constrained}

Since the scenario where all datacenter regions possess an unbounded capacity is not feasible, we also aim to assess the potential reductions under a constrained setting by considering capacity constraints. Adding a simple constraint allows us to gauge the disparity between the ideal and constrained settings.

Figure~\ref{fig:spatial-cap-figs}(b) illustrates the carbon reductions associated with spatial shifting when we assume that all regions worldwide possess identical capacity and are currently operating at 50\% idle capacity. To examine the upper bound of a capacity-constrained setting, we also want to maximize the carbon reductions in this scenario. In pursuing this, the region with the highest carbon-intensity would transfer its workloads to the region with the lowest carbon-intensity. Subsequently, the region with the second-highest carbon-intensity would relocate its tasks to the second-lowest carbon-intensity location, and so forth. Figure~\ref{fig:spatial-cap-figs}(b) shows that at 50\% idle capacity, the carbon reductions from spatial migration decrease to 190 g$\cdot$CO$_2$eq (i.e.,  52\% of the global average emission) as compared to the reduction of 352 g$\cdot$CO$_2$eq when there is no capacity constraint. Of particular interest is that the carbon reductions in Asia remain at 556 g$\cdot$CO$_2$eq. This illustrates the significant proportion of regions with high carbon-intensity located in Asia, and relocating workloads from these regions yields substantial carbon reductions from spatial migration.

Lastly, Figure~\ref{fig:spatial-cap-figs}(c) shows global average reduction (in \%) as the global idle capacity increases. We assume uniform resources across all regions worldwide, and idle capacity refers to the available capacity within each region for accommodating new workloads.
In the most extreme scenario, with idle capacity reaching 99\%, the greenest region has the capability to accommodate all workloads from other regions. As a result, the emission rate is $\sim$16 \emissionunit, which reflects the carbon-intensity of Sweden, the greenest region in our data set. Conversely, when the global idle capacity reaches 0\%, the emission rate levels off at 368.39 \emissionunit, representing the global average carbon-intensity. In this scenario, no workloads can be relocated to other regions, resulting in each region solely managing its own workload. An overall idle capacity of 99\% reduces the global carbon emissions by 95.68\%, and in the less extreme case, a  global idle capacity of 50\% reduces the global emission by 51.5\%. 
While increasing idle capacity helps regions to increase carbon reductions from spatial workload migration, the originating datacenters remain underutilized, which increases operational and non-operational costs such as embodied carbon.

\begin{visionbox}{}
\noindent\emph{\textbf{Key Takeaway.}  In the practical case where datacenter capacity is finite, the carbon reductions from spatial shifting depend on the idle capacity in the system. For a mean utilization of 50\%, the global average carbon reductions drop to 51.5\%, a 1.9$\times$ decrease when compared to our ideal case.} 

\end{visionbox}
\begin{figure}[t]
    \begin{minipage}{0.5\linewidth}
        \includegraphics[width=\linewidth]{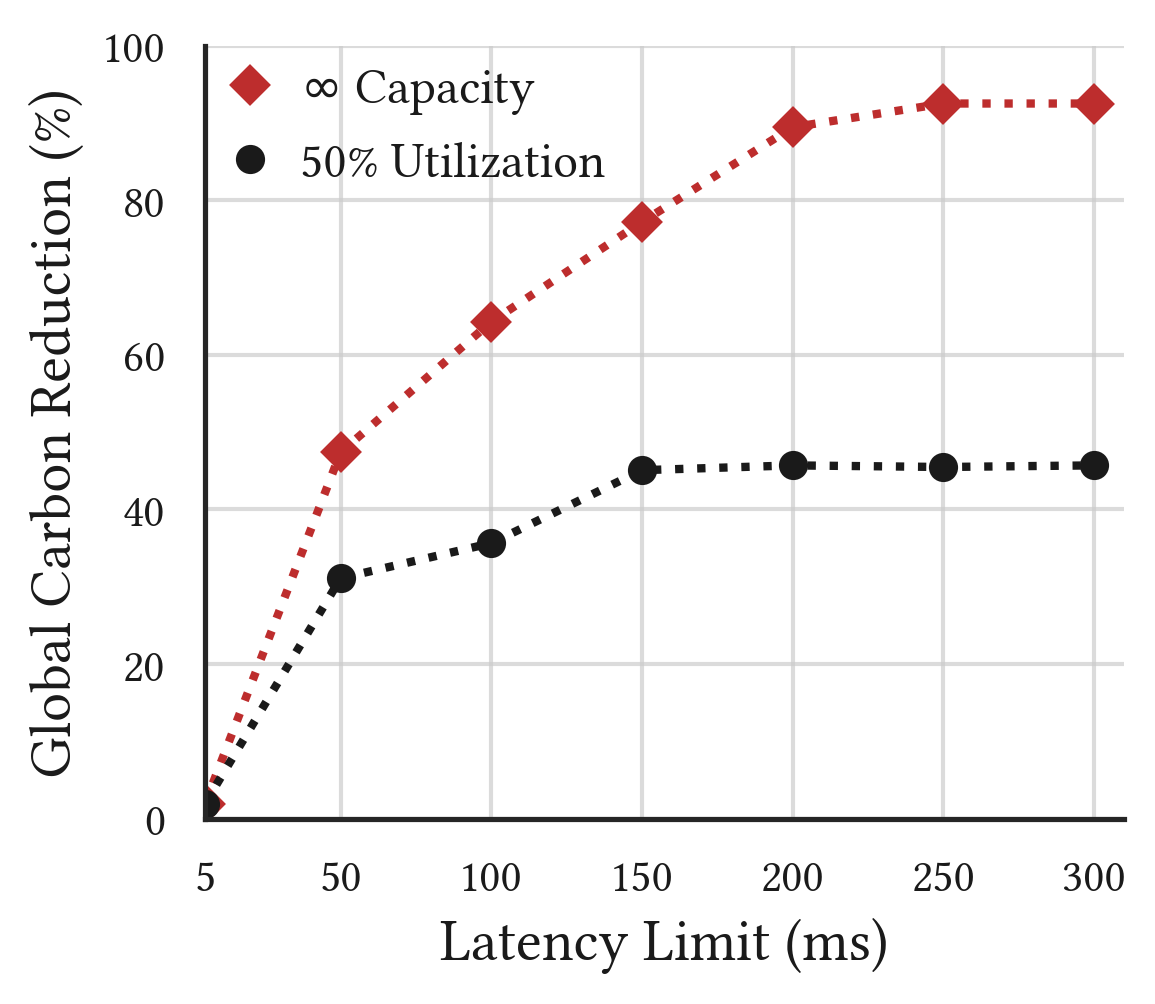}
        \subcaption{Capacity-Latency}
    \end{minipage}\hfill
    \begin{minipage}{0.5\linewidth}
        \includegraphics[width=\linewidth]{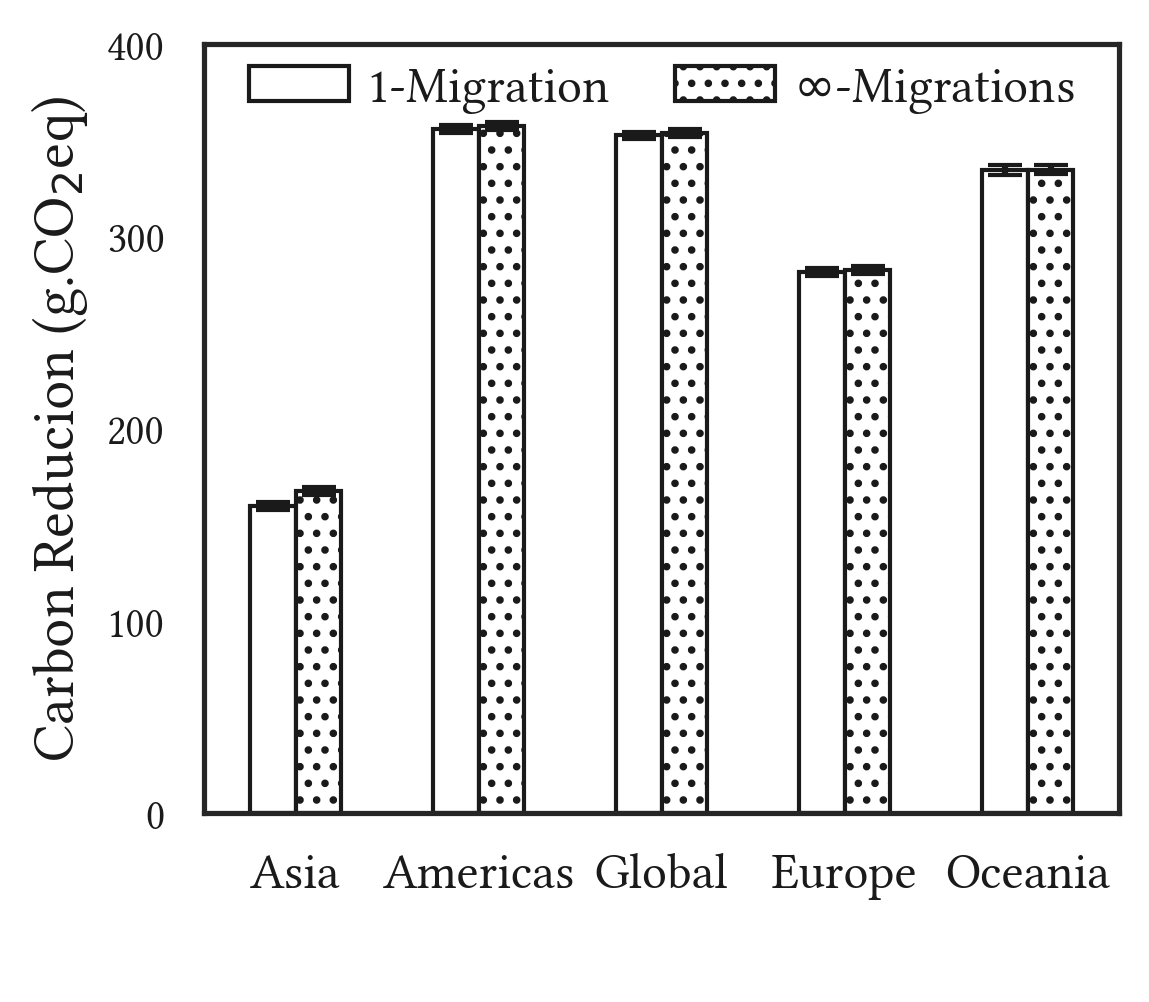}
        \subcaption{Smart Migration}
    \end{minipage}
   \vspace{-0.25cm}
    \caption{\emph{Carbon reduction with capacity and latency constraints and from one and $\infty$-migration policies.}}
    \vspace{-0.35cm}
    \label{fig:latency-and-smart-migration}
\end{figure}

\subsubsection{Carbon Reduction with Capacity and Latency Constrains}
\label{subsec:spatial-latency-constrained}
Apart from the capacity constraints, spatial migration could be restricted by performance constraints such as response time. An ideal workload for spatial migration is the interactive requests originating from web-based services or inference-serving AI/ML systems. These workloads typically do not have any data dependency; thus, their requests can be processed at any location worldwide as long as the user receives the response to the request within a specific time frame. If a request can afford additional latency, it can be migrated to a greener datacenter. 

Figure~\ref{fig:latency-and-smart-migration}(a) shows the global average carbon reduction as a function of latency SLOs. We use the actual latency data for GCP that provides average latency information between two cloud endpoints within GCP~\cite{gcp_latency}. Moreover, we also analyze the ideal and constrained case when the capacity is infinite and 50\% utilized, respectively. We see that with a latency of at least 250ms, all the regions can migrate to the greenest region, and the global carbon emission is reduced by 92.5\% when there is an infinite capacity. However, when the utilization is at 50\%, the global carbon emission is reduced only by 45.7\%. This shows that there is a trade-off between the carbon reductions, latency increase, and datacenter capacity.

\vspace{-0.2cm}
\begin{visionbox}{}
\noindent\emph{\textbf{Key Takeaway.} When latency constraints are added on top of the capacity constraints,
the carbon reduction from spatial shifting decreases further, with the average global carbon reductions dropping from 52\% to 31\% for a 50ms latency constraint and 50\% datacenter utilization.}

\end{visionbox}

\begin{figure}[t]
	\begin{minipage}{0.5\linewidth}
		\includegraphics[width=\linewidth]{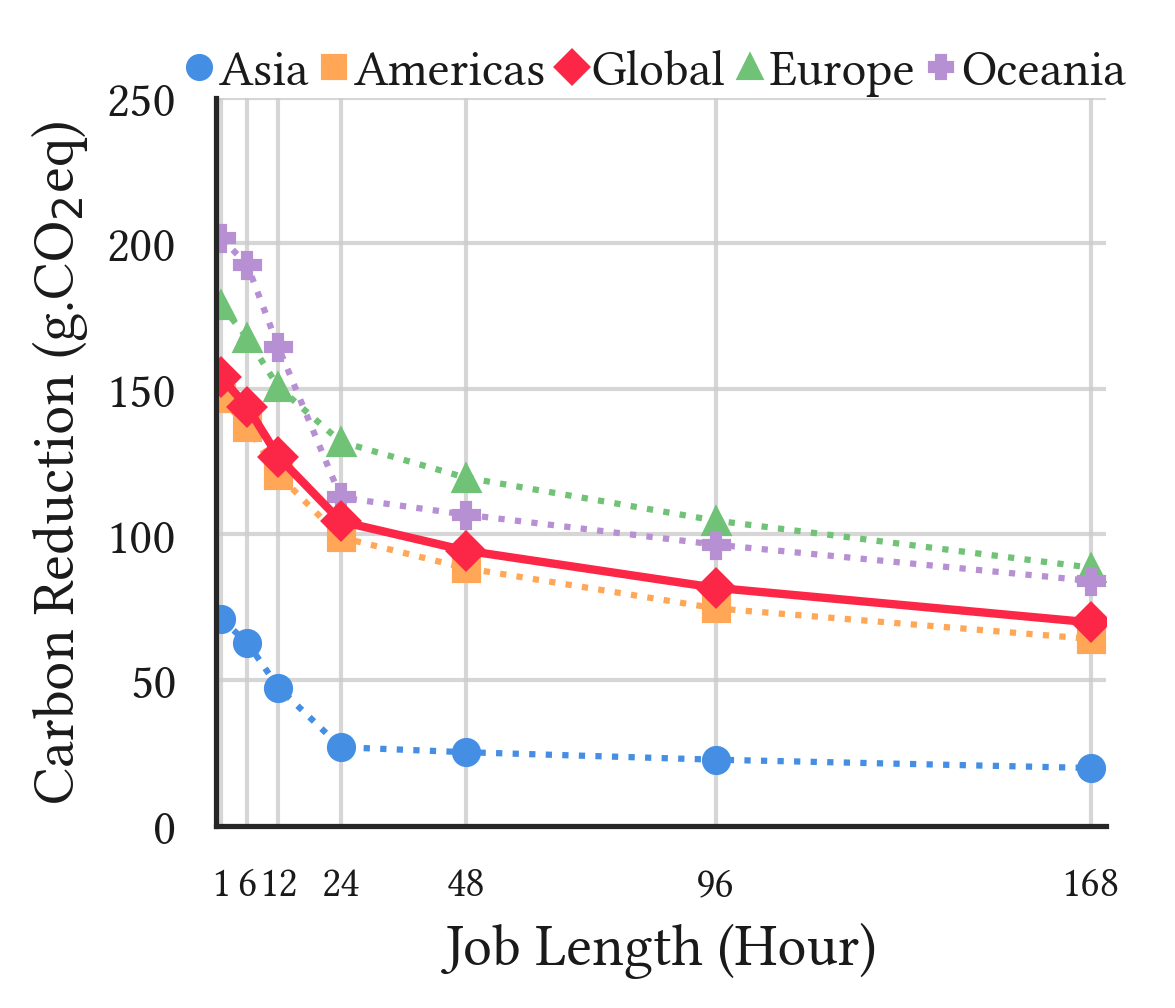}
		\subcaption{One-year slac}
	\end{minipage}\hfill
	\begin{minipage}{0.5\linewidth}
		\includegraphics[width=\linewidth]{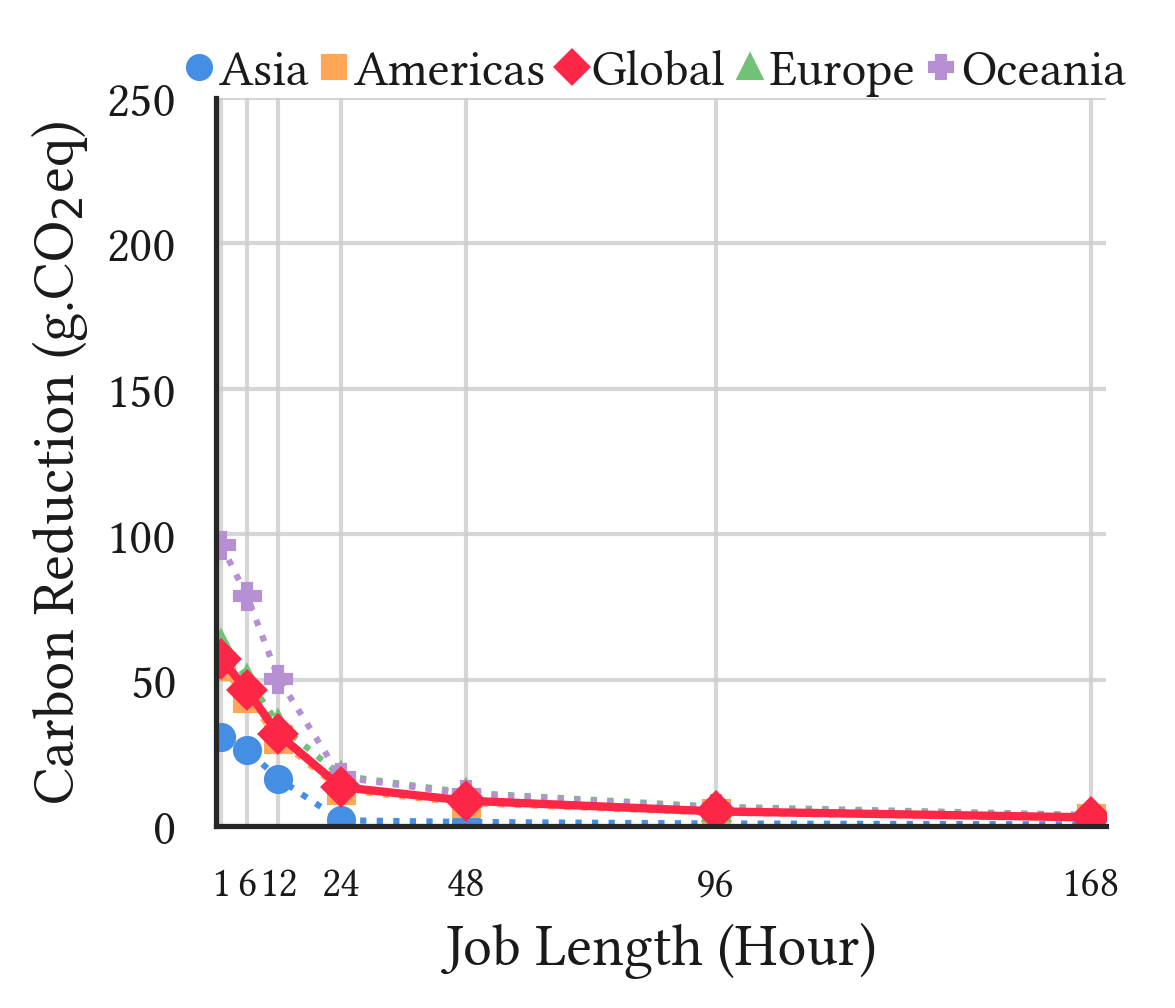}
		\subcaption{24H slack }
	\end{minipage}
	\vspace{-0.25cm}
	\caption{\emph{Carbon reduction from deferrability, normarlized by the job length.}}
	\vspace{-0.35cm}
	\label{fig:temporal-defer}
\end{figure}

\vspace{-0.4 cm}
\subsubsection{Smart Region-hopping Policies}
\label{subsec:policies}

Until now, our default spatial shifting policy has been to migrate once to the greenest available region. We chose this policy because it enables a simple spatial shifting policy based on historical averages which do not vary significantly on a per-year basis. However, as we discussed in \S\ref{sec:introduction}, there is an assumption that the carbon-intensity patterns of various regions vary daily and weekly and thus may overlap or cross. In this case, a more sophisticated migration policy that hops between locations may yield more carbon reductions than a simple one-time migration policy. To evaluate this hypothesis, we devise another spatial region-hopping policy that is clairvoyant, does not incur energy overhead to migrate, and immediately shifts to the destination. We refer to this policy as $\infty$-migrations, since it can migrate an infinite number of times, as compared to the 1-migration policy we have used so far. While we have been exploring spatial migration to the greenest region \emph{globally}, we want to constrain the migration to be within their geographical grouping in this experiment. This is to abstract out the \emph{global} lowest and any strict rankings for all the global traces. 


The results in Figure \ref{fig:latency-and-smart-migration}(b) show that carbon reduction for both policies are relatively similar, with $\infty$-migration yielding only slightly higher carbon reductions but the difference between them being $<10$g$\cdot$CO$_2$eq. Thus, migrating once to the greenest region yields the vast majority of the carbon reductions, and more sophisticated migration approaches that migrate more often are unnecessary. Notably, our $\infty$-migration represents a best-case policy, as it ignores the overhead of migration. Thus, any practical policy outperforming 1-migration would have a very tight upper bound on its carbon reduction. In our current electric grid, this upper bound is less than 10 g$\cdot$CO$_2$eq. As a result, there are limited practical benefits to sophisticated migration policies.
\vspace{-0.2cm}
\begin{visionbox}{}
\noindent\emph{\textbf{Key Takeaway.} A single migration to the greenest cloud region offers the majority of the benefits associated with spatial shifting. Even a clairvoyant policy that overlooks migration overheads reduces, at most, an additional 10 g$\cdot$CO$_2$eq carbon emissions compared to the single migration policy.}
\end{visionbox}

\subsection{Temporal Shifting}
\label{sec:temporal}
\begin{figure}[t]
    \begin{minipage}{0.5\linewidth}
        \includegraphics[width=\linewidth]{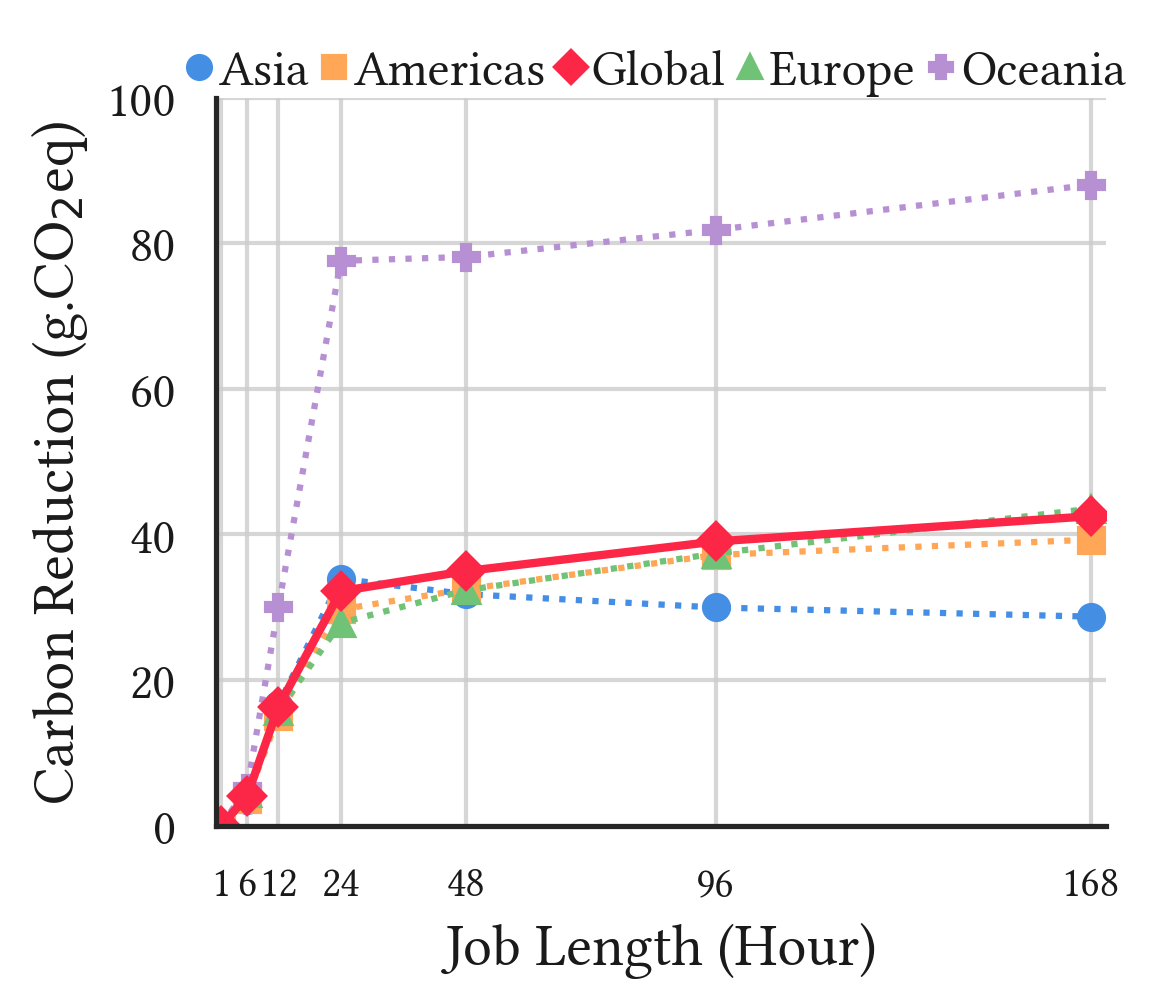}
        \subcaption{One-year slack}
    \end{minipage}\hfill
    \begin{minipage}{0.5\linewidth}
        \includegraphics[width=\linewidth]{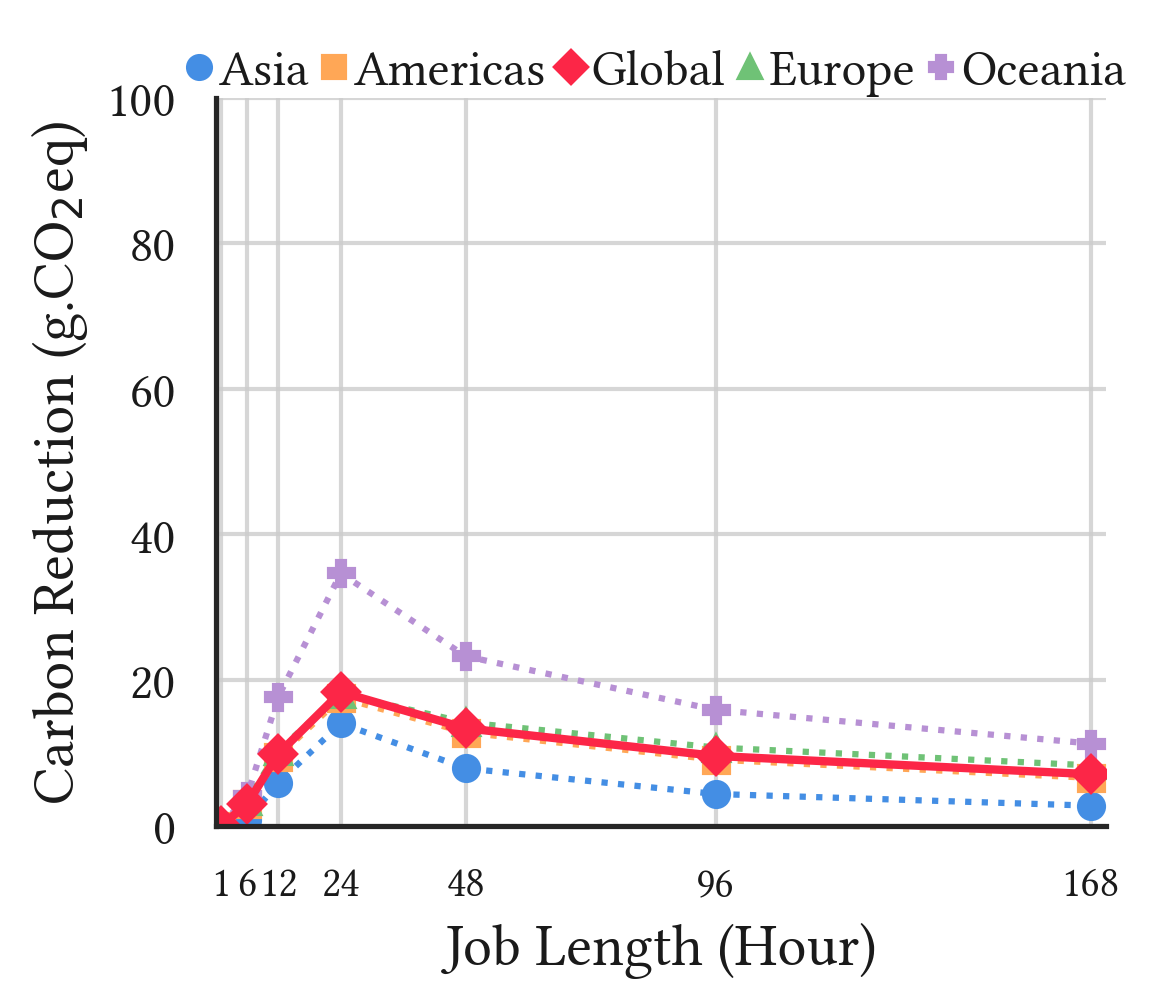}
        \subcaption{24H slack}
    \end{minipage}
    \vspace{-0.4cm}
    \caption{\emph{Carbon reduction from interruptibility, normarlized by the job length.}}  
    \vspace{-0.45cm}
    \label{fig:temporal-interrupt}
\end{figure}

In this section, we quantify the upper limits of temporal workload shifting by considering a scenario where workloads have a perfect knowledge of the future carbon-intensity for the whole year. We refer to  this as the one-year slack case. With this level of flexibility, a specific job can choose the most favorable time slot to start execution, therefore achieving the lowest carbon emissions in a region. We then also analyze the carbon reductions for a more realistic choice of slack which is a 24-hour slack, for it provides a sufficiently generous buffer within real-world settings~\cite{pradeep:sc20}. We explore two main dimensions of temporal flexibility of a workload, namely  deferrability and interruptibility. We also analyze how the combination of both deferrability and interruptibilty affects the carbon reductions.

\subsubsection{\textbf{Effects of Deferrability}}
\label{subsec:deferrability}
In this analysis, we examine the carbon reductions from deferrability for both one-year and 24-hour slack settings. The carbon reductions are relative to a non-deferrable job of the same length running in the same region. The datapoints are averages across all start times over a year. Figure \ref{fig:temporal-defer}(a) shows the theoretical upper bound on carbon reduction from deferrability, normalized by the job length, for each job length. 
The figure illustrates that carbon reductions per unit job decrease from 154 to 70 g$\cdot$CO$_2$eq as the job length increases from 1 hour to 168 hours, representing a reduction of 41.8\% and 19\% relative to average global emissions, respectively.

This is because as the job length increases, the job has to start occupying the “peaks” in the carbon trace despite being given the slack, while jobs that are <24 hours can utilize the slack to find the lowest valleys in the carbon trace. In a practical scenario, Figure \ref{fig:temporal-defer}(b) shows that the global carbon reductions decrease from 57 g$\cdot$CO$_2$eq to 3 g$\cdot$CO$_2$eq. This is due to the relatively low fixed slack duration (24h) in comparison to the job lengths. A 168-hour job with a 24-hour slack will have few opportunities to find periods with low carbon-intensity compared to smaller jobs.

\begin{figure}[t]
    \begin{minipage}{0.5\linewidth}
        \includegraphics[width=\linewidth]{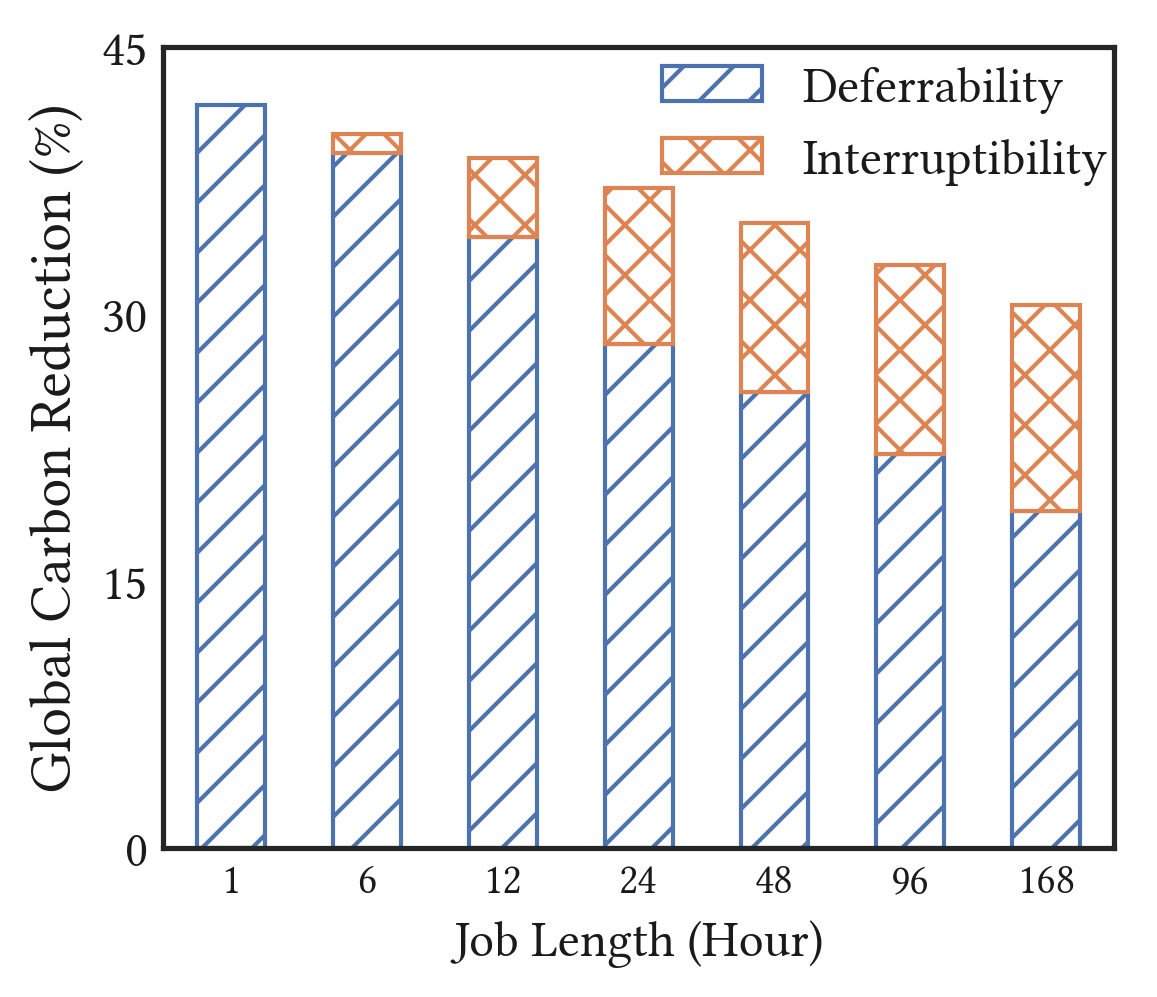}
        \subcaption{One-year slack}
    \end{minipage}\hfill
    \begin{minipage}{0.5\linewidth}
        \includegraphics[width=\linewidth]{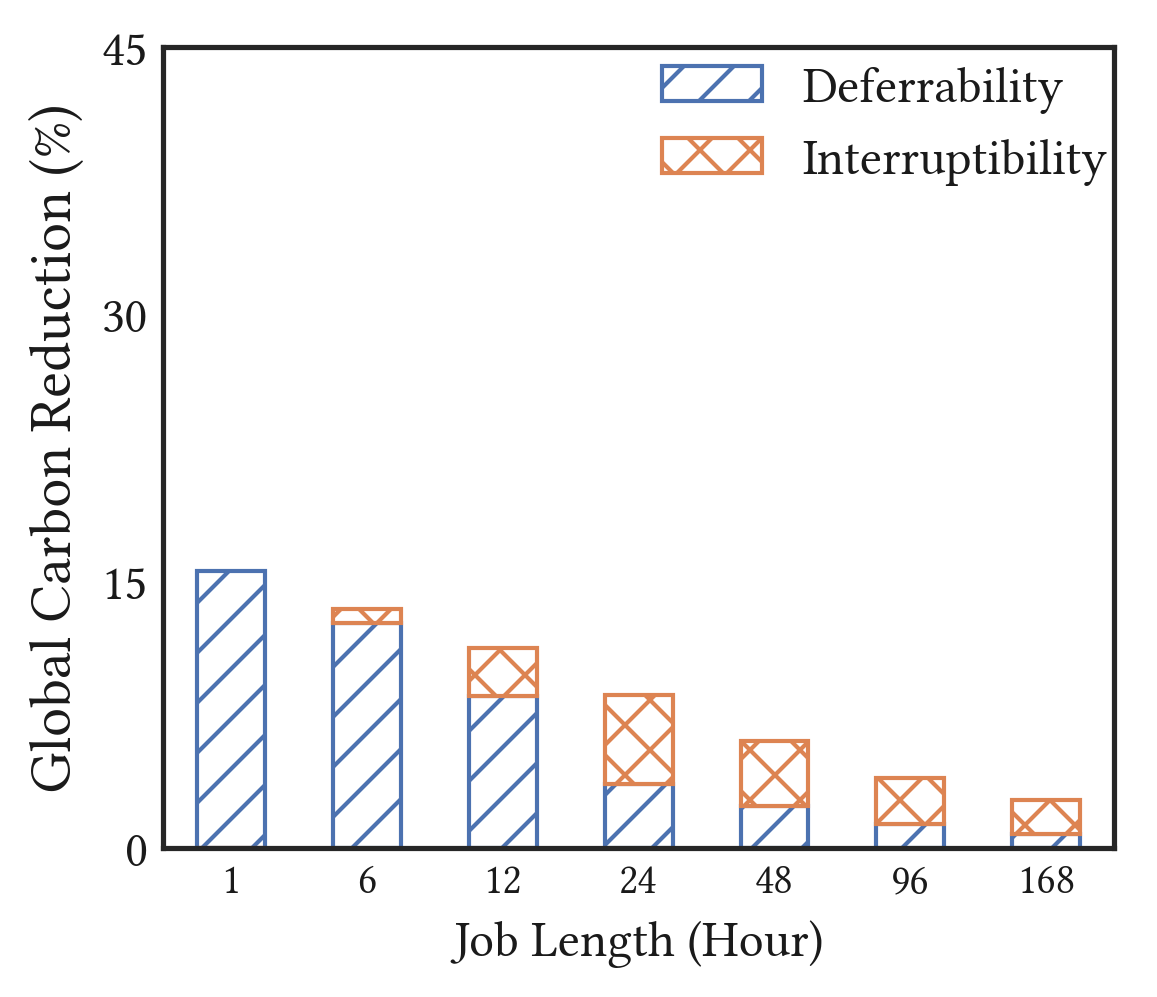}
        \subcaption{24H slack}
    \end{minipage}
    \vspace{-0.4cm}
    \caption{\emph{Average carbon reduction from deferrabiity and interruptibility across all job lengths from all regions, with respect to the global average.}}  
    \vspace{-0.6 cm}
    \label{fig:temporal-defer-interrupt-stacked}
\end{figure}
\begin{figure*}[t]
    \begin{minipage}{0.24\linewidth}
         \includegraphics[width=\linewidth]{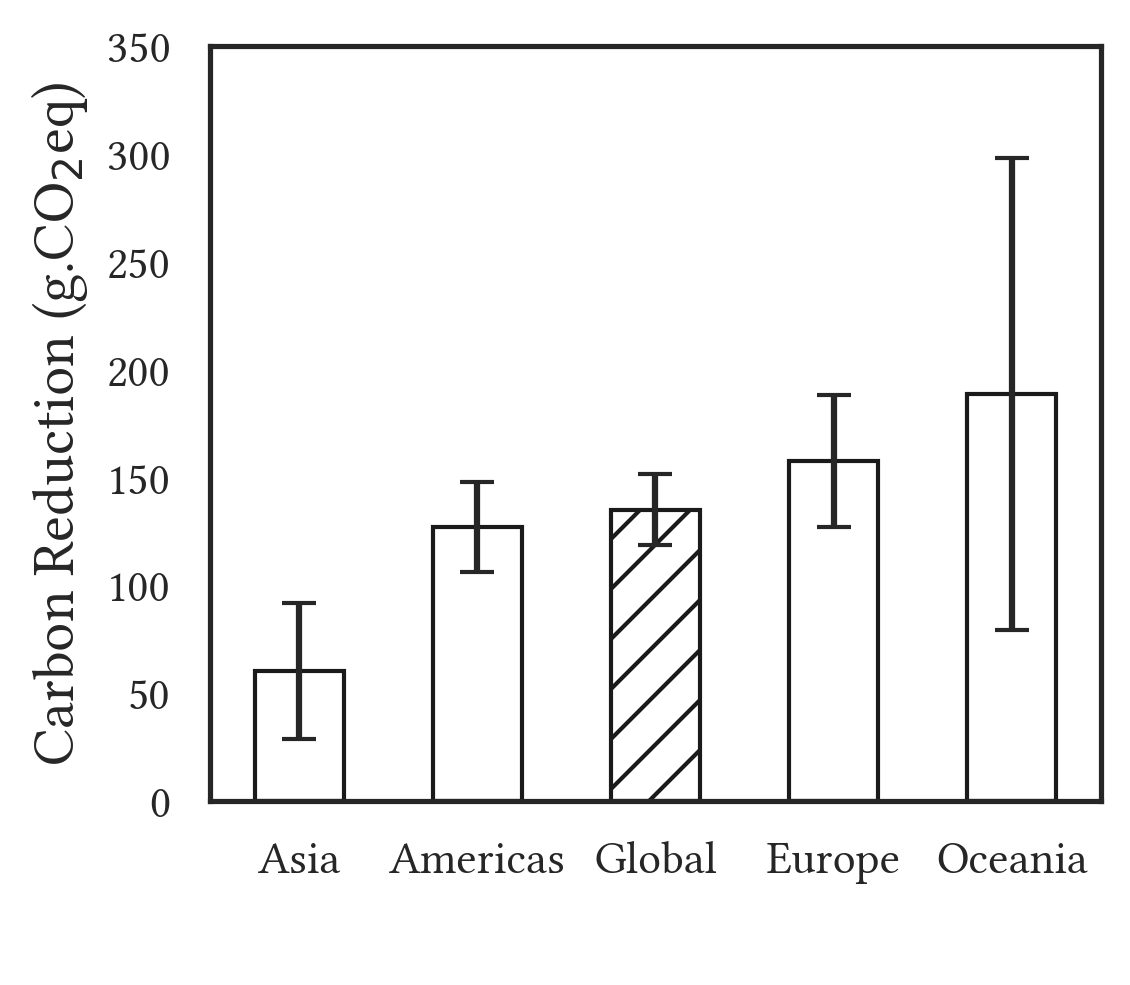}

        \subcaption{Equally Distributed}
    \end{minipage}\hfill
    \begin{minipage}{0.24\linewidth}
        \includegraphics[width=\linewidth]{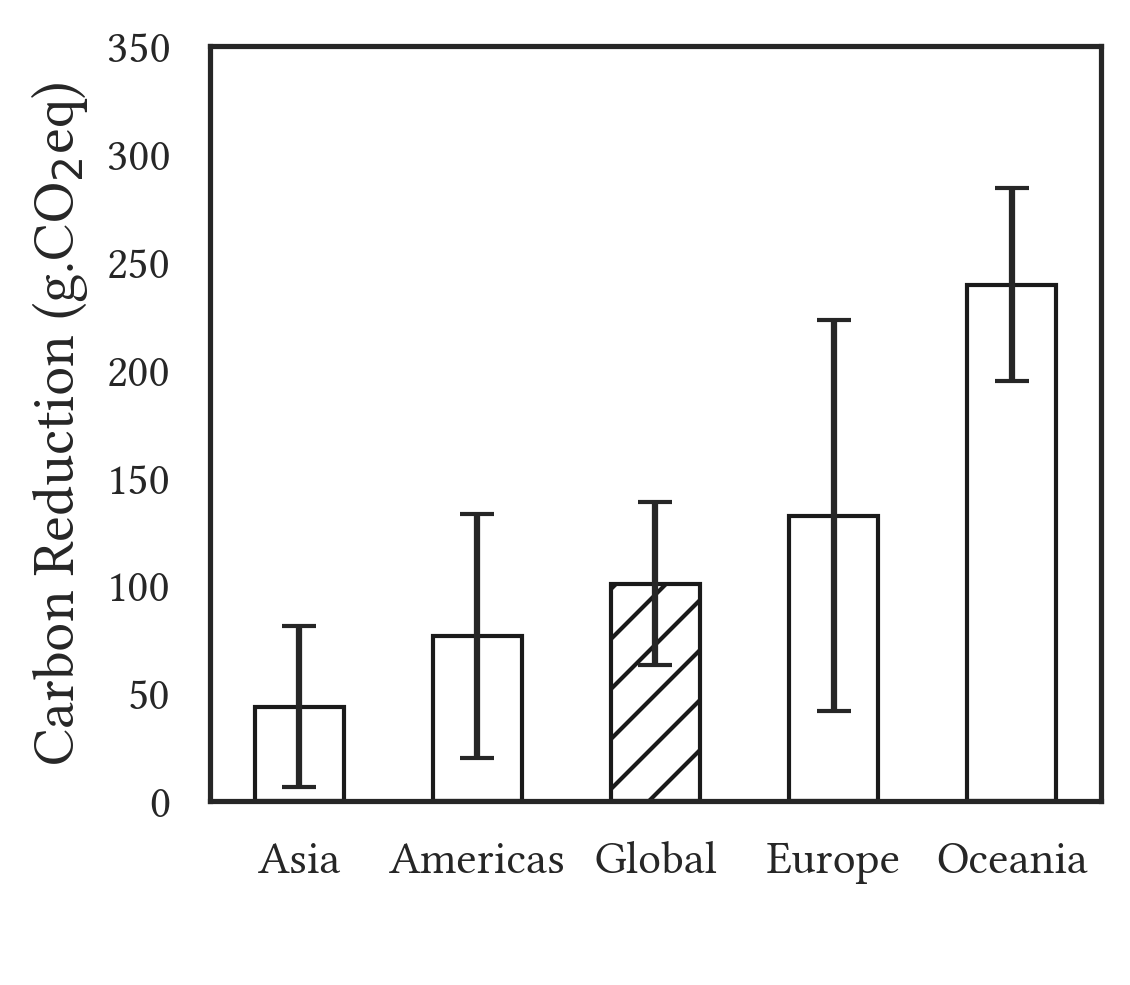}

        \subcaption{Azure}
    \end{minipage}\hfill
    \begin{minipage}{0.24\linewidth}
        \includegraphics[width=\linewidth]{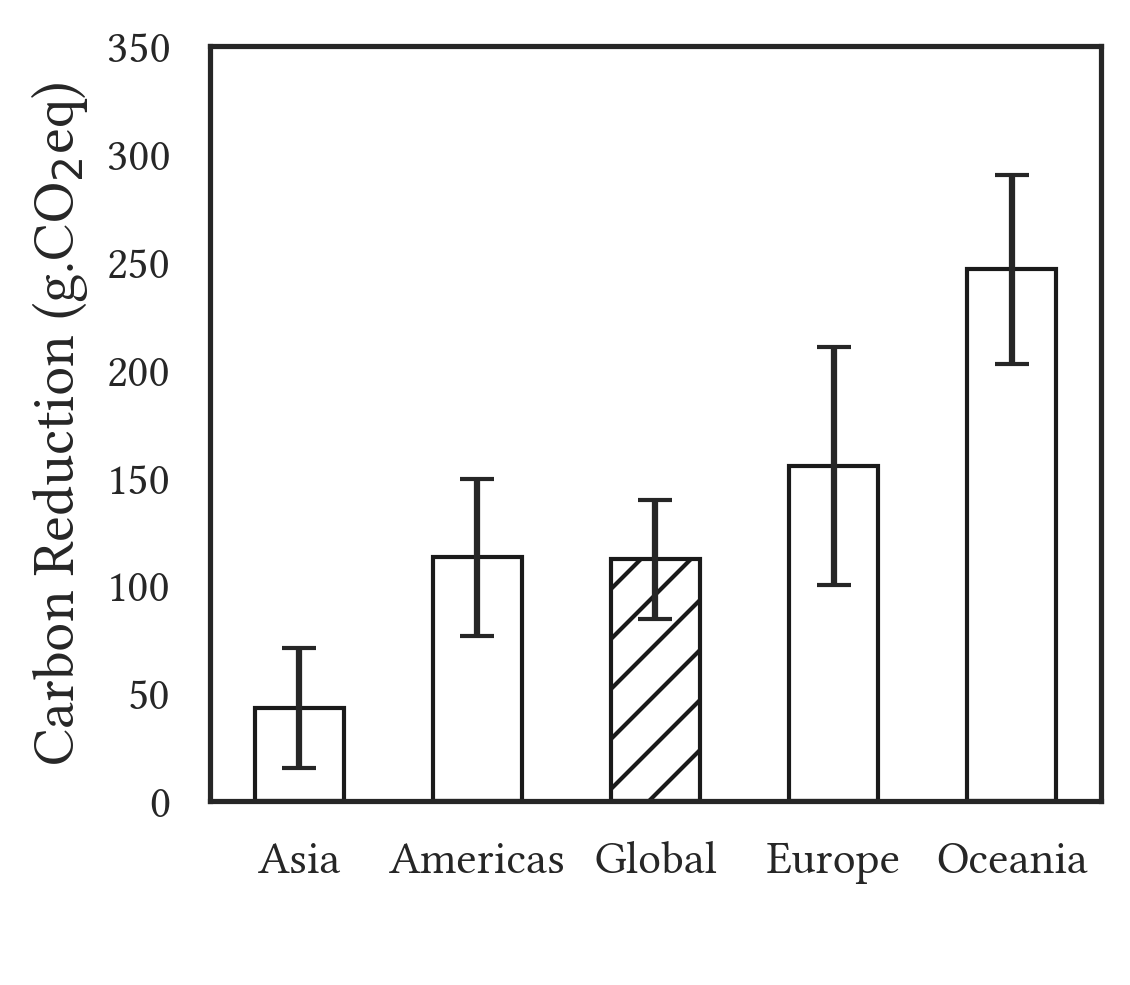}
        \subcaption{GCP}
    \end{minipage}
    \begin{minipage}{0.24\linewidth}
        \includegraphics[width=\linewidth]{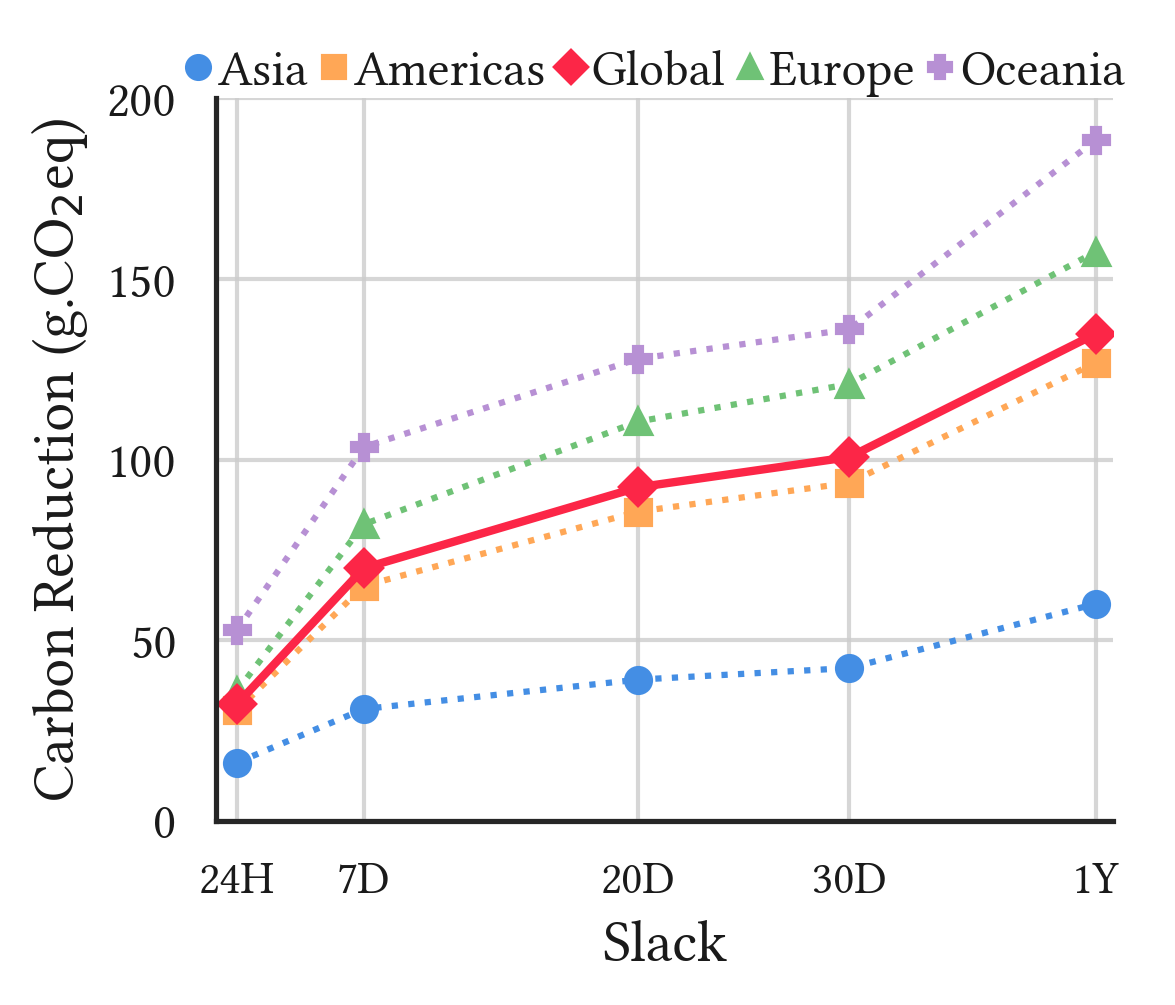}
        \subcaption{Different Slacks}
    \end{minipage}
 
    \caption{\emph{Estimation of global and regional carbon reductions through the temporal shifting of workloads under different distributions of job lengths with one-year slack, as well as upper bound of carbon reductions for different choices of slacks}}

    \label{fig:temporal-real_trace}
\end{figure*}

\subsubsection{Effects of Interruptibility}
\label{subsec:interuptibility}

In addition to deferring a job’s start time, the ability to interrupt an already running job when energy’s carbon-intensity increases is another dimension of temporal flexibility. Interruptibility enables schedulers to pause jobs when energy’s carbon-intensity is high and resume them when it is low. Figure \ref{fig:temporal-interrupt}(a) shows the upper bound of additional carbon reductions from interruptibility, normalized by the job length, for each job length if they are interruptible and deferrable. Here, the carbon reductions from interruptibility increase from 0 to 43 g$\cdot$CO$_2$eq per job length as the job length increases. The carbon reductions from interruptibility is 0 for a 1-hour job because the smallest granularity of a batch job in our setup is 1 hour, so it cannot be interrupted. Besides a 1-hour job, carbon reductions per job length from interruptibility increase mainly because long jobs can now be broken down into smaller jobs and run in low-carbon time slots throughout the year. While long jobs can exploit the effect of interruptibilty in an ideal case Figure \ref{fig:temporal-interrupt}(b) shows that the carbon reduction reaches its peak at 24-hour job lengths at 18.4 g$\cdot$CO$_2$eq and gradually decreases as the job length increases. The 24-hour jobs exhibit the highest carbon reductions because the 24-hour slack aligns with the job’s length, allowing the interruptibility policy to select very low-intensity periods from two carbon-intensity “valleys” within the available 48-hour time window. Smaller jobs, however, gain lower carbon reductions as most potential reductions are gained simply by deferring them. Jobs longer than 24 hours also gain lower carbon reductions as they need to occupy some “peaks” in the carbon trace even when interruptible.

\subsubsection{Combined Deferrability and Interuptibility}
\label{subsec:deferrability_interuptibility}

Figure \ref{fig:temporal-defer-interrupt-stacked} shows the breakdown of global average carbon reductions from deferrability and interruptibility across all regions. Figure \ref{fig:temporal-defer-interrupt-stacked}(a) shows that in the ideal setting, the benefits of deferrability decrease as the job length increases. However, interruptibility helps leverage carbon reductions for long jobs. For 168-hour job, with just deferrability, the carbon saving is only at 19.0\%, but interruptibilty helps improve the carbon reductions by another 11\%. In the practical scenario, however, even with interruptibility, the carbon reductions remain small for long jobs. Figure \ref{fig:temporal-defer-interrupt-stacked}(b) shows that a 168-hour job only gains 3\% of carbon reductions even if the job is deferrable and interruptible. This is because there are limited low-carbon regions for long jobs to run; hence, some parts of the job will run on a high-carbon period.

\begin{visionbox}{}{
\noindent\emph{\textbf{Key Takeaways. }
Overall, when considering carbon reductions from both deferrability and interruptibility, the shorter the job length, the higher the carbon reduction per unit job. Short jobs see up to 154 g$\cdot$CO$_2$eq of carbon reduction per unit job length, a 42\% carbon reduction relative to  global average emissions.}}

\end{visionbox}

\vspace{0.2cm}

\subsubsection{Savings with Equal Distribution}
\label{subsec:temporal-inf-slack}

Figure \ref{fig:temporal-real_trace}(a) shows the upper limit of temporal carbon reductions for the one-year slack case with equally distribution job lengths (see \S\ref{sec:methodology}). With a one-year slack, the carbon reduction for global grouping is 135 g$\cdot$CO$_2$eq, 37\% relative to  average global emissions. For other geographical groupings, the average temporal carbon reductions range from 60 g$\cdot$CO$_2$eq in Asia to 189 g$\cdot$CO$_2$eq in Oceania, 16\% and 51\% relative to the average global emissions, respectively. Next, we analyze how the distribution of job lengths impacts real cloud workloads.

\begin{figure*}[t]
    \begin{minipage}{0.24\linewidth}
         \includegraphics[width=\linewidth]{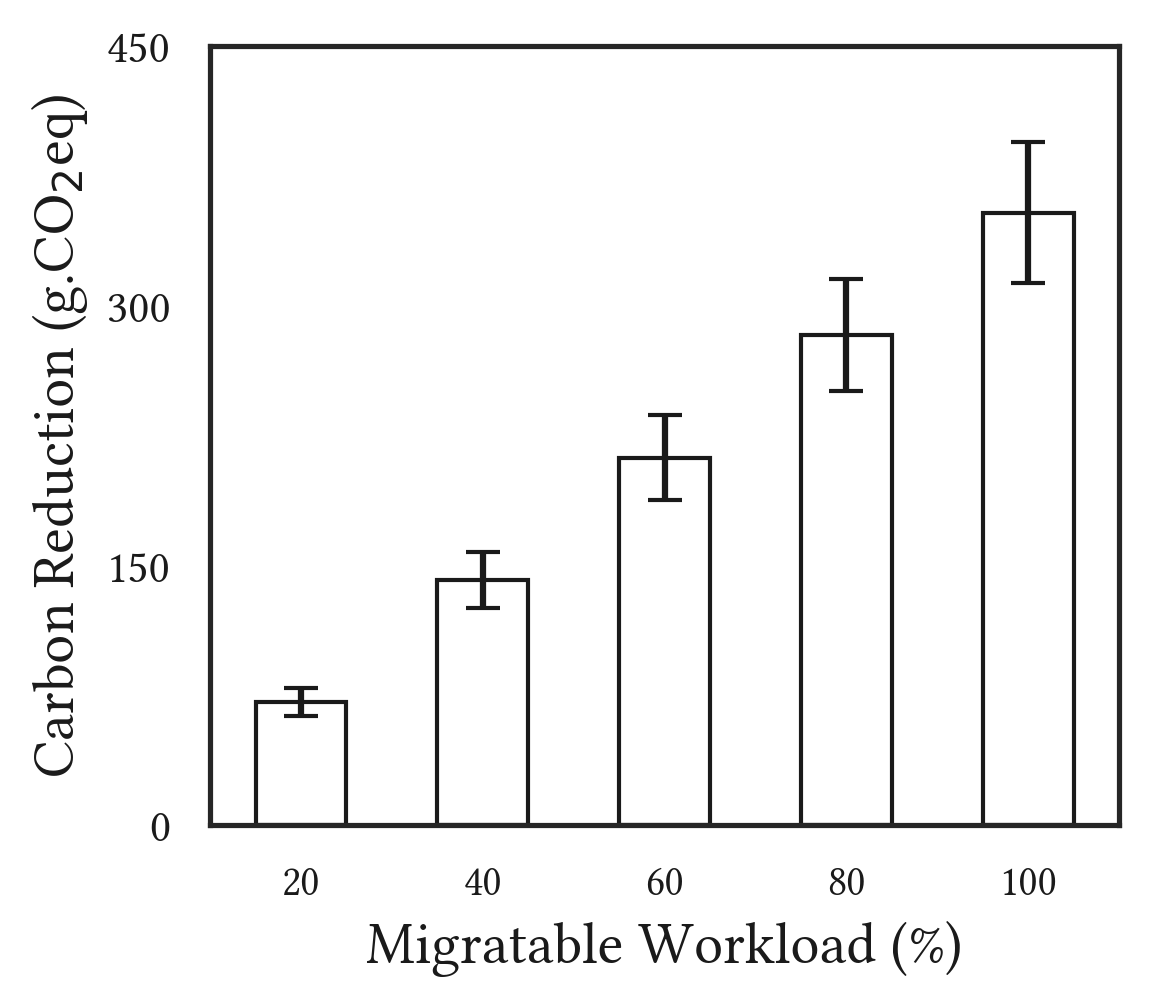}
        \subcaption{Mixed Workload}
    \end{minipage}\hfill
    \begin{minipage}{0.24\linewidth}
        \includegraphics[width=\linewidth]{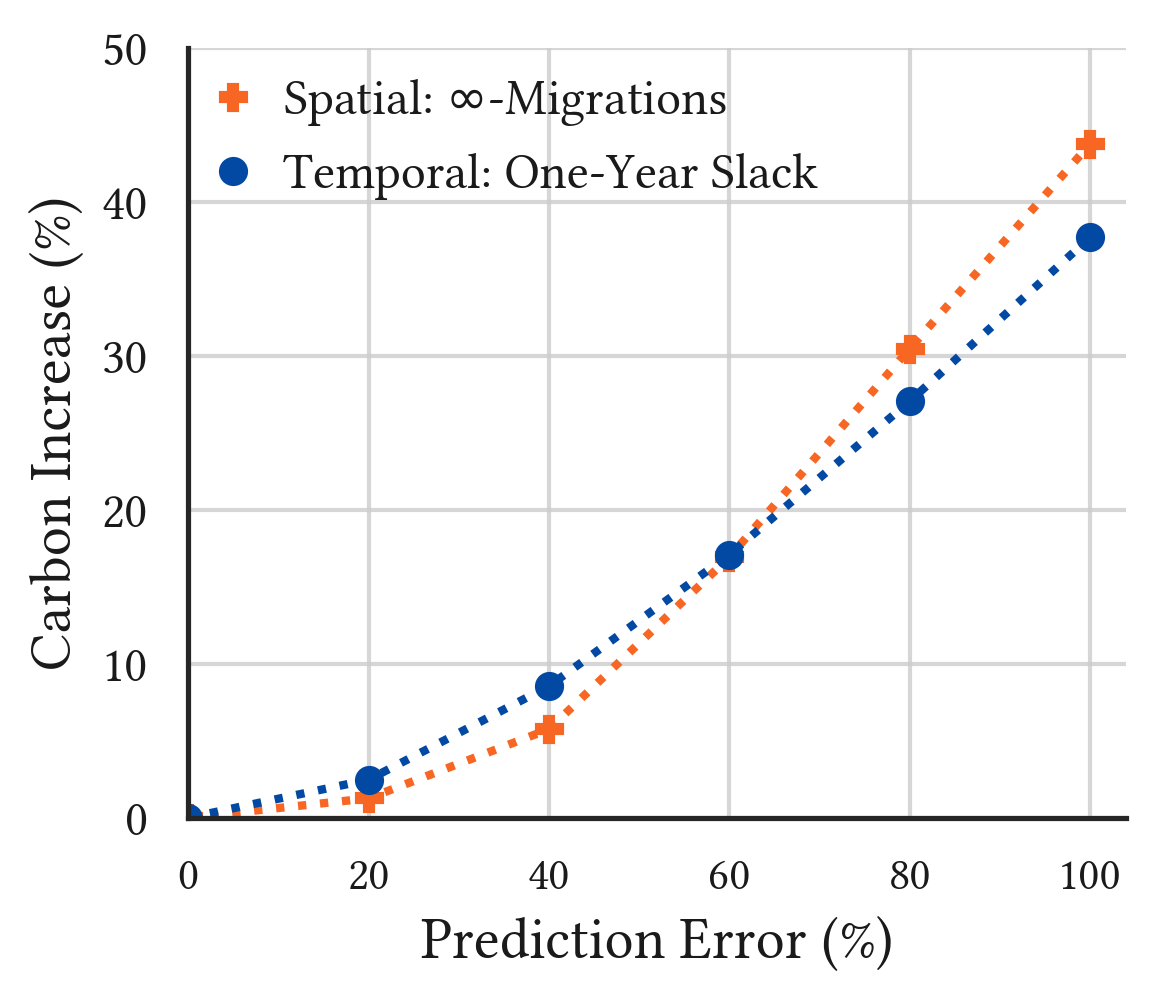}
        \subcaption{Prediction Error}
    \end{minipage}
    \begin{minipage}{0.24\linewidth}
        \includegraphics[width=\linewidth]{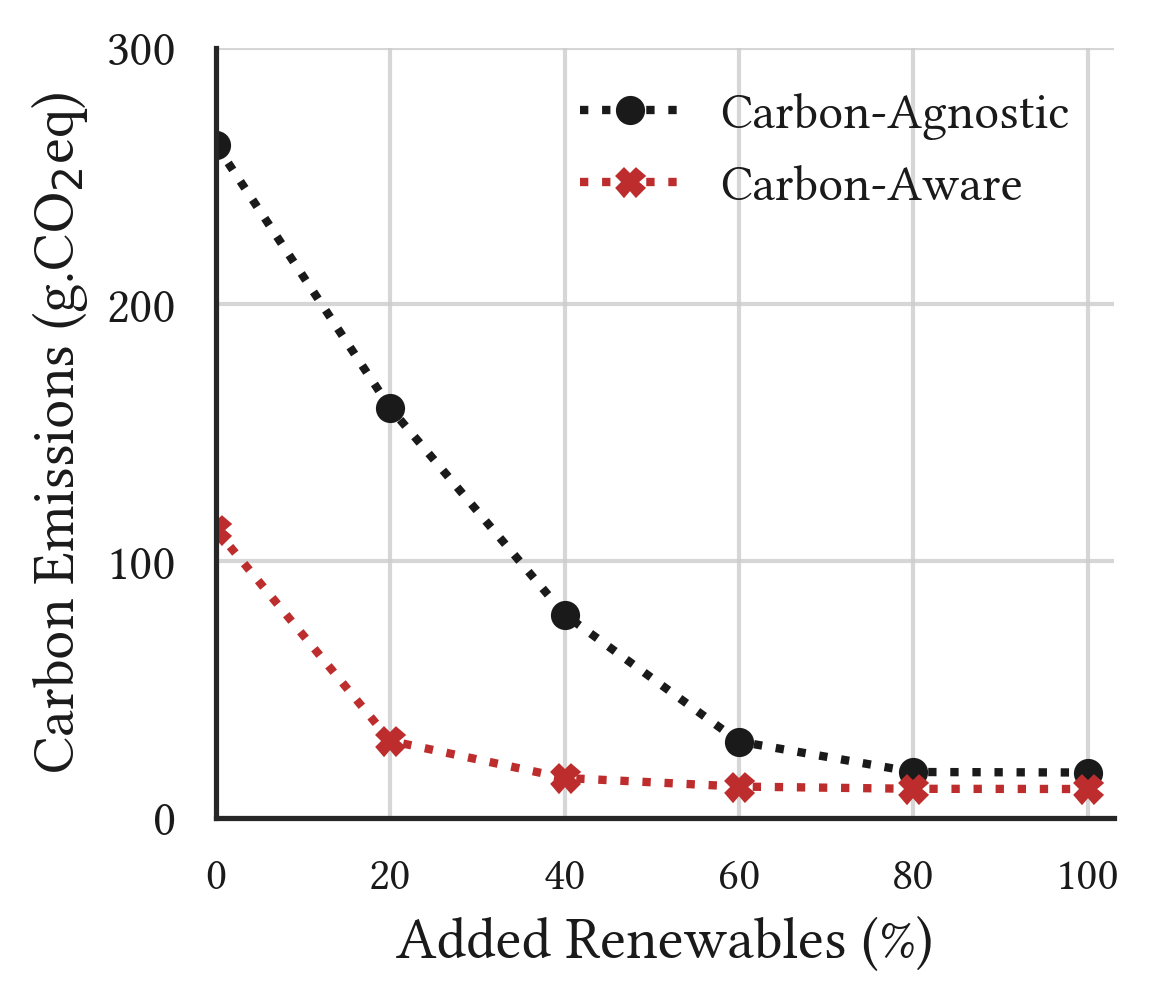}

        \subcaption{Temporal: California (US)}
    \end{minipage}\hfill
    \begin{minipage}{0.24\linewidth}
        \includegraphics[width=\linewidth]{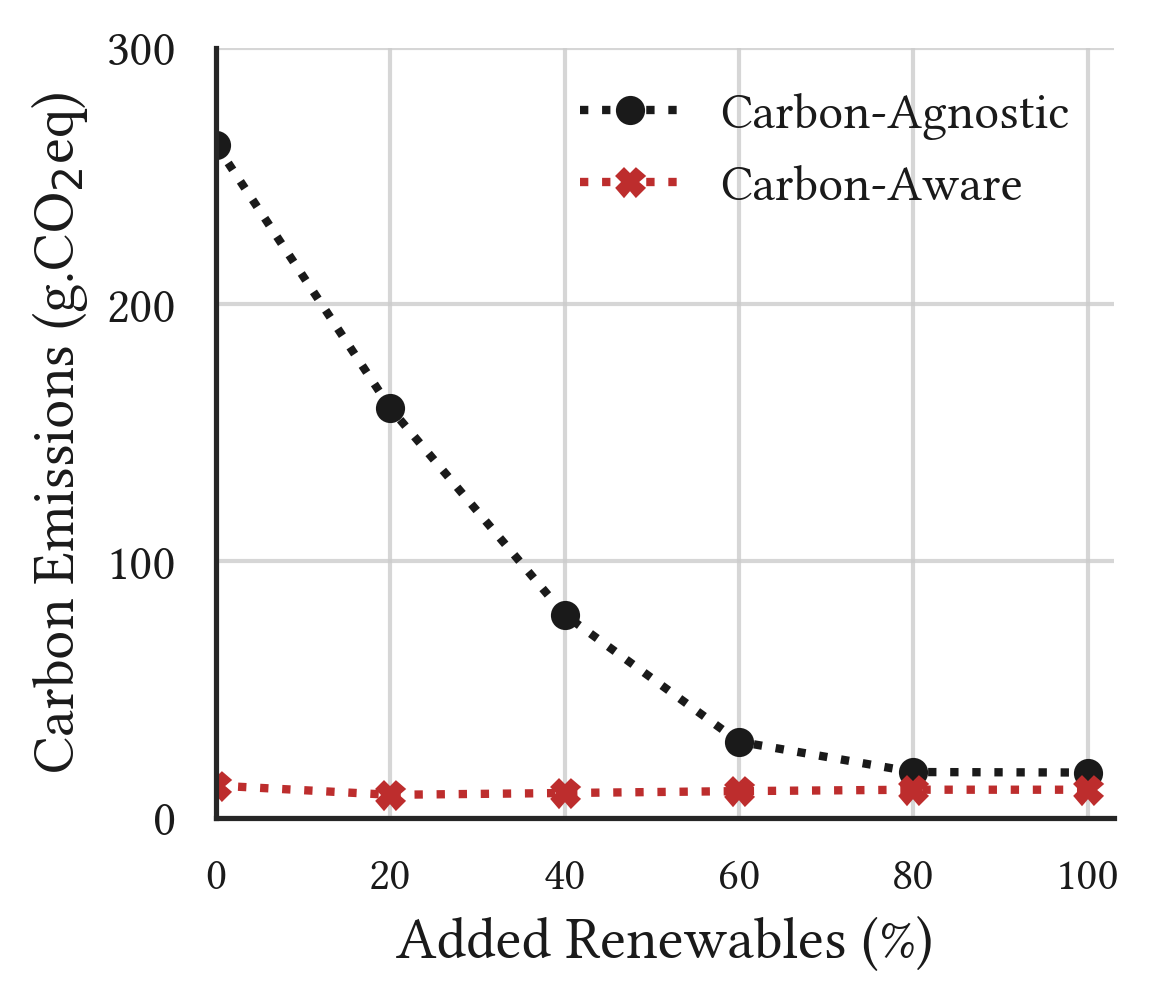}
        \subcaption{Spatial: California (US)}
    \end{minipage}
    \caption{\emph{Impact on carbon reductions and emissions when a only a portion of workloads is migratable, when there is a prediction error, and when a sample region, California, becomes greener.}} 
    \label{fig:what-if-scenarios}
\end{figure*}

\subsubsection{Savings with Cloud Workloads Distribution}
\label{subsec:temporal-cloud}

Figure \ref{fig:temporal-real_trace}(b) and Figure \ref{fig:temporal-real_trace}(c) show the upper limits of cloud workload carbon reductions for the one-year slack case weighted by the workload distributions in Azure~\cite{azure} and Google\cite{borg-next-gen} traces. While the average carbon reduction reductions in the equally weighted workload is 135 g$\cdot$CO$_2$eq, the average carbon reductions for Azure and Google are 100 g$\cdot$CO$_2$eq and 112 g$\cdot$CO$_2$eq, respectively (27\% and 30\% relative to the average global emissions). The Azure and Google traces have a much higher percentage of long jobs ($>$48hrs), so even with the large one-year slack and interruptibility, the carbon reductions per unit job are still low, as discussed in Section~\ref{subsec:deferrability_interuptibility}. In particular, 1\% of the very long-running jobs ($>$1 week) account for 90\% of resource utilization and energy consumption in the Google trace~\cite{borg-next-gen}.

In general, the carbon reductions are high in regions with high variations in their carbon-intensity. The high variations indicate high renewable penetration. For example, Oceania has a higher percentage of renewables and carbon reductions ranging from 246 g$\cdot$CO$_2$eq to 189 g$\cdot$CO$_2$eq 67\% and 51\% relative to the average global emissions. for different job distributions. On the contrary, the carbon-intensity in Asia's regions is low as the regions mostly rely on fossil fuels. Hence, the carbon reductions in Asia have a small range of 43 g$\cdot$CO$_2$eq to 60 g$\cdot$CO$_2$eq. While our analysis provides the upper bound on ideal temporal shifting for different job distributions, the actual carbon reductions are likely to be much less due to less slack for long jobs, resource constraints that prevent running many jobs during low carbon periods, and uncertain carbon-intensity forecasts.  We next examine the effects of slack on temporal carbon savings.

\subsubsection{Savings with Varying Slacks}
\label{subsec:temporal-pratical-slack}

Figure \ref{fig:temporal-real_trace}(d) shows temporal carbon savings for different slacks. With slack varying from 24 hours to one year, the carbon reductions vary from 31 g$\cdot$CO$_2$eq to 127 g$\cdot$CO$_2$eq, 8.4\% and 34\% relative to the average global emissions. The large range of carbon reductions translates to a large disparity in carbon reduction from temporal shifting between the ideal and constrained settings. These results also suggest that temporal carbon savings exhibit sub-linear growth. Notably, beyond a 7-day timeframe, no substantial increases in savings are observed in any of the regions.
From 24 hours to one year, the slack increases by 365$\times$ while the carbon reductions only increase by 3.1$\times$. This implies that harnessing carbon savings through temporal shifting in real-world scheduling settings would require an exceptionally large amount of slack, an impractical proposition for most environments.

\vspace{-0.2cm}
\begin{visionbox}{}
\noindent\emph{\textbf{Key Takeaways.}
For real cloud workloads, the carbon reductions from temporal shifting are limited to 112 g$\cdot$CO$_2$eq, and depend heavily upon factors like long job lengths, substantial slack, absence of resource constraints, and access to future knowledge of carbon-intensity.}
\end{visionbox}

\subsection{System Design Implications}
\label{sec:design-implication}
Below, we discuss the system design implications of our analysis of spatial and temporal workload scheduling. 

\subsubsection{Spatial Shifting}
\label{subsec:spatial_summary}
 
Overall, the workloads that are migratable will benefit from spatial shifting as they can run in green regions to reduce carbon emissions. Our findings also reveal a significant correlation between idle capacity and the benefits of spatial shifting: for every 1\% increase in global idle capacity, there is a corresponding $\sim$1\% reduction in global average emissions, equivalent to around 3.68 \emissionunit. 
 
Nevertheless, this ultimately results in increased idleness in high intensity regions, and in over-provisioning in low intensity ones, which could have adverse implications on datacenters' embodied carbon costs, primarily due to sub-optimal utilization of resources. 



In addition, from a stability standpoint, most datacenters operate at low utilization to ensure they can service their customers' demand~\cite{huang2022metastable}.
This suggests datacenters in low carbon intensity regions should focus on optimizing their idle efficiency, e.g., by transitioning servers to low power states when idle. In contrast, high carbon intensity regions can reduce their emissions through energy efficiency, for instance, with measures like Dynamic Voltage and Frequency Scaling (DVFS) that curtail energy consumption in response to workload intensity. Additionally, spatial migration over shorter distances through e.g., edge computing, may prove more advantageous than relying on cloud resources because it can better exploit local renewable availability.

\subsubsection{Temporal Shifting}
\label{subsec:temporal_summary}
The workloads that benefit most from temporal shifting include mostly short jobs. Our findings highlight the importance of aligning job executions with periods of low carbon-intensity. 
However, the most substantial benefits of temporal shifting are enjoyed by small jobs, which decrease significantly as the duration of the workload approaches full daily cycles. 
Small jobs offer greater flexibility in determining their start times, and they can further leverage pause and resume techniques. 
That is, despite larger jobs typically having more slack, the only viable option for operators is to use suspend and resume techniques to align the execution of large jobs with low carbon-intensity periods.
When formulating carbon-efficient policies that use temporal shifting to enhance carbon efficiency, it is crucial to prioritize small jobs that can be scheduled during minimal valleys in a region's intensity trace. 
Additionally, many jobs can be broken down into smaller interconnected processing components, effectively constituting a workflow of several smaller jobs that can benefit from temporal shifting. Leveraging such workflow information and integrating it into carbon-aware orchestration, jointly with information from the local cloud or edge, is a viable approach to reduce the carbon emissions of long-running jobs.

\vspace{-0.2cm}

\section{What-If Scenarios}
\label{sec:what-if}
This section explore multiple ``what-if'' scenarios for spatiotemporal workload shifting. In particular, we explore the effect of i) mixed batch and interactive workloads, ii) carbon forecasting errors, iii) increasing renewable penetration, and iv) combining spatial and temporal shifting.

\vspace{-0.2cm}

\subsection{Mixed Workload}

We assume mixed workloads consist of two workload classes---migratable (batch) and non-migratable (interactive)---and present the carbon reductions with respect to the proportion of the migratable workload. The non-migratable workloads are executed in the arrival region at the arrival time, while the migratable workloads are migrated and executed in the region with the lowest carbon-intensity at the arrival time. 

Figure~\ref{fig:what-if-scenarios}(a) shows that as the percentage of the migratable workloads increases, the carbon reduction also increases, as the workloads can exploit spatial flexibility and migrate to the low-carbon region at the time period. Examining the effects of mixed workloads on carbon reduction helps us understand carbon reductions in a real-world context. In some scenarios, roughly 30\% of VMs are interactive, such that the workloads have strict SLOs and are not migratable~\cite{cortez2017resource}. 

\begin{figure}[t]
	\centering
	\includegraphics[width=1.0\linewidth]{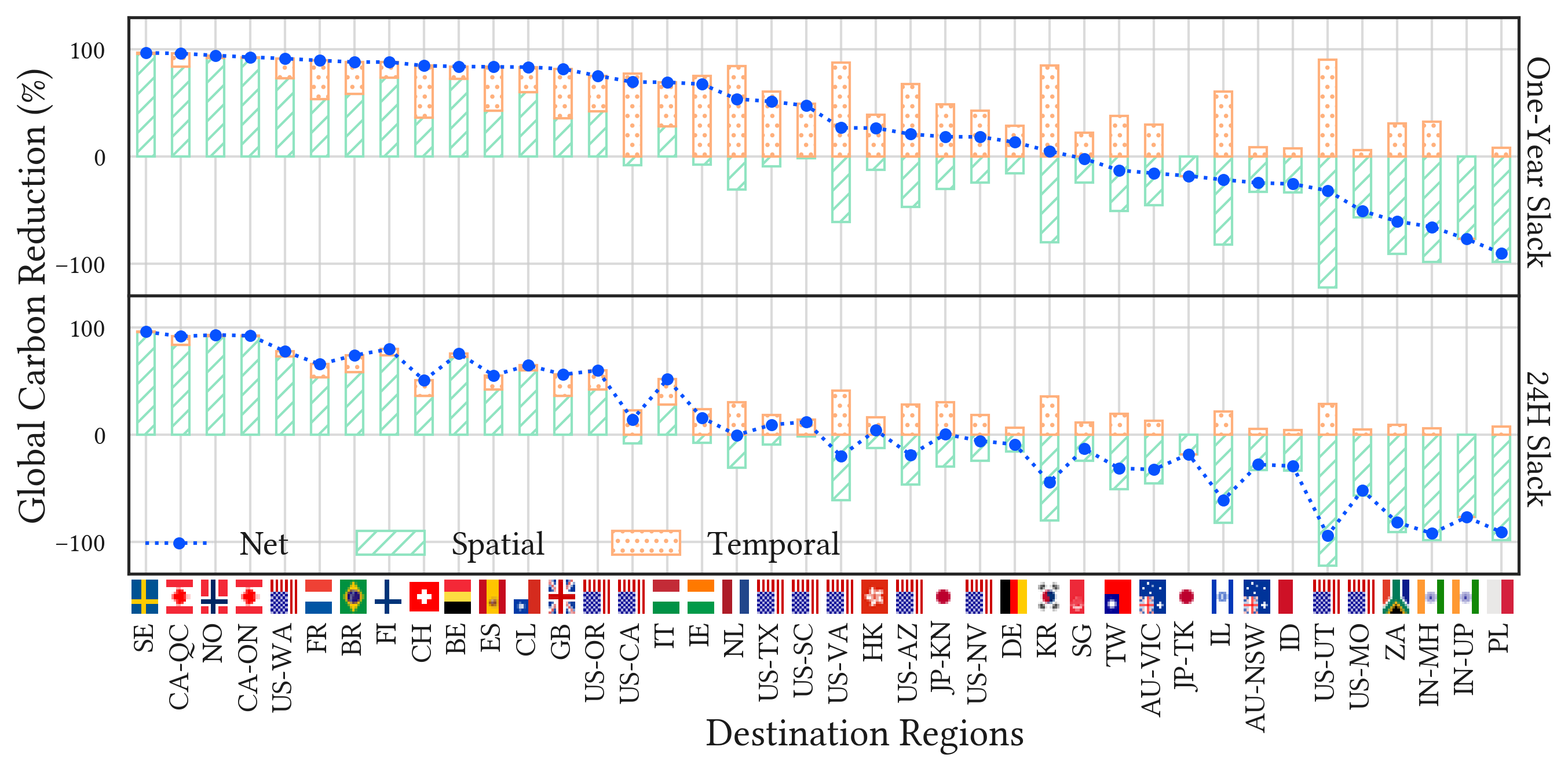}
	\caption[]{\emph{Breakdown of spatial and temporal carbon reduction with net reductions for one-year and 24h slack. Flags indicate destination regions where jobs from all regions migrate to.}}
	\label{fig:spatial_temporal_combined}
\end{figure}

\subsection{Prediction Error}
We also consider the effect of forecast error. Here, we add different amounts of forecast error, in this case from a uniformly random distribution, to assess the impact of such errors. In particular, for each value of prediction error, the workload is scheduled temporally and spatially based on an inaccurate carbon intensity value.

In temporal scheduling, we consider all job lengths at least 1 hour (see~\S\ref{sec:methodology}) with a one-year slack. To calculate the carbon increase in temporal shifting based on the prediction error, we first calculate the carbon emissions when the temporal scheduling is done with the error-free carbon trace, and then compute the difference when using the inaccurate forecast.
In spatial scheduling, we consider the $\infty$-migrations policies based on the error-free and error-added trace. For each prediction error value, we determine the region with the lowest carbon intensity of that time interval. We then account for the actual carbon intensity from the error-free traces of that region in that time slot. We also account for the lowest carbon intensity for each time interval with the traces that have no error.

In both temporal and spatial scheduling, the carbon increase from the prediction error is the difference between the carbon emissions scheduled based on the error-added trace and those scheduled based on an error-free carbon trace.
Figure~\ref{fig:what-if-scenarios}(b) shows that the carbon increase from temporal and spatial scheduling is $\sim$10-12\% when the prediction error is 50\%. For context, prior work on CarbonCast has a mean absolute percentage error (MAPE) up to 14\%~\cite{maji2022carboncast}, which implies  a $\sim$3\% increase in carbon emissions due to forecasting errors in practical settings.

\subsection{Increasing Renewable Penetration}

Figure~\ref{fig:what-if-scenarios}(c)-(d) shows the carbon emissions from carbon-agnostic and carbon-aware scheduling. In both temporal and spatial scheduling, carbon-agnostic scheduling executes jobs immediately in the arrival region without having any slack or migration. For temporal carbon-aware scheduling in Figure~\ref{fig:what-if-scenarios}(c), the workload has a one-year slack and for spatial carbon-aware scheduling in Figure~\ref{fig:what-if-scenarios}(d), the workload uses the $\infty$-migrations policy discussed in~\S\ref{subsec:policies}. We see that in both temporal and spatial scheduling, as a region becomes ``greener,'' the benefits of carbon-aware spatiotemporal workload shifting will increase. This can be seen by how the emissions from carbon-aware scheduling  decrease in Figure \ref{fig:what-if-scenarios}(c) and (d).
However, as the percentage of renewables grows, the region's average carbon intensity  also decreases, causing carbon-agnostic scheduling to yield lower emissions. This implies that the benefits of carbon-aware scheduling diminish when compared to carbon-agnostic methods as the world's energy supply becomes greener.

\subsection{Spatial and Temporal Shifting Combined}
\label{subsec:spatial_and_temporal_combined}

Finally, we examine the combined effect of spatial and temporal shifting, quantifying their limit in ideal and practical settings. 
The approach allows jobs to not only migrate to different regions but also leverage deferrability to schedule their execution during optimal periods within the destination region. 
We consider two scenarios: the ideal case, where jobs can defer execution by up to a year, and the constrained case, where the deferral window is limited to 24 hours. 

Figure~\ref{fig:spatial_temporal_combined} shows that, in general, regardless of the choice of slack, carbon reduction from spatial workload shifting dictates a decision on whether a workload should migrate to a particular region. That is, spatial reductions dominate temporal reductions regardless of their sign (positive or negative). For example, migrating workloads to Sweden (SE), Ontario (CA-ON), or Belgium (BE) results in high net carbon reduction because migrating to one of these regions yields high carbon reductions from spatial shifting, even if these regions have low variability and therefore low temporal carbon reductions. 
Conversely, migrating workloads to the Netherlands (NL), South Korea (KR), or Utah (US-UT) yields low to negative net gains in carbon reductions (i.e. incurring more carbon emissions as compared to running the job locally) even when these regions have high temporal reductions. This is because these regions have negative gains in carbon reductions from spatial migration. Of course, there are some exceptions where regions have low spatial gains and high temporal gains and yet result in relatively high net carbon reductions, such as California (US-CA) and Virginia (US-VA). However, in general, reductions from spatial migration when migrating to a region influence the net gains in carbon reductions.

\vspace{-0.2cm}
\begin{visionbox}{}
\noindent\emph{\textbf{Key Takeaways.}
When combining spatial and temporal
shifting, savings from spatial migration dominate the overall savings, with limited additional benefits from performing temporal shifting. In addition, the benefits of carbon-aware spatiotemporal optimizations decline as regions adopt higher levels of renewable energy.}
\end{visionbox}

\section{Related Work}
\label{sec:related-work}
There has been much prior work that proposes carbon-aware spatiotemporal workload shifting policies to reduce computing's carbon footprint~\cite{ecovisor,dean-carbon,wait-awhile, radovanovic2021carbonaware,hanafy2023carbonscaler, igsc2023casper, maji2023bringing,dodge2022measuring,radovanovic2021carbonaware, zhou2013carbon}. Our work is related to this work in using these spatiotemporal shifting policies to evaluate an upper bound on carbon savings in different scenarios.

\noindent \textbf{Spatial Shifting.}  
Prior work~\cite{dean-carbon,dodge2022measuring} discusses how the choice of region to execute the workload impacts the carbon emissions. Prior work~\cite{ igsc2023casper, maji2023bringing} also proposes specific region selection techniques to reduce carbon emissions. To quantify an upper limit on spatial workload shifting, our work also examines region selection policies for workload execution and evaluates optimal region selection policy, namely $\infty$-migrations policy.

\noindent \textbf{{Temporal Shifting.}}
Prior work also exploits temporal flexibility by delaying workload execution~\cite{radovanovic2021carbonaware, carbonexplorer}, using suspend-resume techniques~\cite{wait-awhile}, or scaling the workload~\cite{hanafy2023carbonscaler, ecovisor}. Our work uses similar variations of deferrable and interruptible temporal workload shifting policies discussed in~\cite{wait-awhile}.

Other prior work explores other aspects of carbon reduction~\cite{maji2022carboncast, hanafy2023war}. CarbonCast is a machine-learning model that provides multi-day forecasts of the grid’s carbon-intensity with a mean absolute percentage error (MAPE) of 4.80–13.93\% across six regions~\cite{maji2022carboncast}. While our work quantifies an upper limit on carbon reduction by assuming perfect knowledge of carbon-intensity, in a practical setting a low-error forecasting model would be required for carbon reduction to approach the upper limit. Further, there is also prior work on the 
tradeoff between carbon-efficiency and other metrics, such as energy-efficiency and cost~\cite{zhou2013carbon,hanafy2023war}. Our work differs in that we focus narrowly on quantifying the upper limit on carbon reduction, and do not consider energy-efficiency and cost.









\section{Conclusions}
\label{sec:conclusion}
In this paper, we conducted an empirical analysis of the benefits and limitations of spatiotemporal workload shifting in the cloud. Our results have several important implications. 

Most importantly, our results show that although there is the potential for some significant carbon savings from spatiotemporal workload shifting, the benefits are often limited in practice. For temporal shifting, these limits derive from a lack of variability in carbon intensity at many locations. In addition, the locations with low variability -- where temporal shifting is least effective --tend to be those with the highest absolute carbon emissions -- where reducing carbon emissions is most important. Likewise, locations with significant variability tend to have low average carbon emissions, and thus adapting to such variations does not yield significant savings.  For spatial shifting, resource constraints will likely limit much of the, potentially significant, carbon savings in practice by preventing most jobs from migrating to the lowest carbon regions.  In addition, migration overheads may also prevent many large jobs that consume significant resources and energy, i.e., from processing large datasets, from benefiting from migration.  

Of course, as the grid becomes greener our results may change.  For example, as more locations adopt renewables, their carbon variability will increase and the global average carbon will decrease.  This will increase both the relative and absolute benefit of temporal shifting, since even the highest carbon regions will have variations in their carbon intensity due to renewables.  A greener grid will also elevate the importance of more sophisticated spatial shifting policies, as there are likely to be more frequent overlaps in the carbon intensity profiles of different nearby locations. 
 
As the grid integrates more intermittent solar and wind to lower its emissions to lower its emissions, it will need more flexible capacity that can dynamically vary its energy consumption to offset variations in these sources.  Thus, rather than adapting to how carbon emissions currently vary in the grid, cloud platforms might be more effective in supporting grid's operations so it can increase the penetration of renewable energy.  Our future work will quantify the potential of cloud platforms in supporting such goals.

\section*{Acknowledgements}
\label{sec:acknowledgement}
We thank the EuroSys reviewers and our shepherd, Shadi Noghabi, for their valuable comments, which improved the quality of this paper. We thank Electricity Maps for access to their carbon-intensity datasets. This research was supported by NSF grants 2213636, 2105494, 2021693, 2020888, 2213636, DOE grant DE-EE0010143, and by VMware.

\bibliographystyle{ACM-Reference-Format}
\bibliography{paper}

\clearpage
\appendix

\appendix
\section{Artifact Appendix} 
\subsection{Abstract}

This artifact is created with the goal of quantifying the limitations of carbon-aware temporal and spatial workload shifting in the cloud. The artifact contains multiple experiments that quantify an upper bound on both the ideal and practical benefits of carbon-aware spatiotemporal workload shifting for a wide range of computing workloads. We conduct a detailed trace-driven analysis to understand the benefits and limitations of spatiotemporal workload scheduling for computing workloads with different characteristics, e.g., job duration, deadlines, and SLOs, based on hourly variations in energy's carbon-intensity over three years across 123 distinct regions, which encompass most major datacenter sites.

\vspace{-0.3cm}
\subsection{Description \& Requirements}
\subsubsection{How to access}
The artifact is available at Zenodo \url{https://zenodo.org/records/10790855} as well as GitHub \url{https://github.com/umassos/decarbonization-potential}. The repository contains more detailed descriptions of the setup and required dataset for each experiment.

\vspace{-0.2cm}
\subsubsection{Hardware dependencies} None
\vspace{-0.2cm}
\subsubsection{Software dependencies}
This artifact has been tested on Ubuntu 20.04 LTS using Python 3.8. The required Python modules are pandas, numpy, scikit-learn, matplotlib and seaborn.
Assuming the dependencies can be successfully installed, we expect the artifact to also work on other Unix-based systems that support Python 3.8+.
\vspace{-0.25cm}
\subsubsection{Benchmarks}
\textit{Carbon Intensity Data}: We collected carbon intensity traces for \regioncount different geographical regions worldwide from 2020 to 2022 using the Electricity Maps web API\footnote{\url{https://www.electricitymaps.com/data-portal}}. Each trace reports the region's average carbon intensity, measured in grams of carbon dioxide equivalent per kilowatt-hour (\carbonunit) hourly. The \regioncount regions are listed in the \url{global_modules/static_files/name_stats.csv}.
\\
\\
\textit{Network Latency}: To simulate the intra-datacenter network latency, we use the Google Cloud Inter-Region latency measurements collected by the AT\&T Center for Virtualization from the Southern Methodist University~\footnote{\url{https://lookerstudio.google.com/s/tN5DUuNd_28}} for the May'2023 period.
\textit{Workload Trace}: We use the workload trace from Azure~\cite{azure} and Google~\cite{google-trace}. The links to the specific workload traces are also provided in the GitHub repository. %
\subsection{Setup}

We generated a \emph{requirements.txt} file for the required Python packages. We suggest users to create a Python virtual environment\footnote{\url{https://docs.python.org/3/library/venv.html}}, update the pip command, and install the required modules inside of this virtual environment. To install the requirements, run \emph{pip install -r requirements.txt}

\subsection{Evaluation workflow}

Before running all the spatiotemporal experiments, we first pre-process the raw carbon intensity and latency data. The scripts for pre-processing the raw carbon intensity and latency data can be found in the \url{process_raw_data} directory. The processed data are stored as .csv files in the \url{shared_data} directory and are used across experiments except otherwise noted.

\subsubsection{Major Claims}

\begin{itemize}
    \item \textit{(C1): Depending on the set of regions in the carbon intensity dataset, the results may differ.}
    \item \textit{(C2): Depending on the year covered by the carbon intensity data, the results may differ.}
    \item \textit{(C3): Depending on the workload trace, the results may differ.}
\end{itemize}

\subsubsection{Experiments}
The directories whose names start with \textit{sim\_} are the directories with different groups of simulations. The sub-directories in each parent directory are individual simulations, which are to be run inside its own directory. 
~\\
\textbf{Experiment (E1): [Mean and Daily CV]:} This experiment generates Figure 3(a), illustrating the yearly mean and average daily coefficient of variation (CV) for carbon intensity across all regions for 2022. This experiment is in \url{sim_trace_analysis/mean_and_cv} directory, and the \url{calculate_mean_and_cv.py} script is run before plotting the results using \url{plot_mean_and_cv.py}.
~\\\\
\textbf{Experiment (E2): [Change Over Time]:} This experiment generates Figure 3(b), showing the change in the region's yearly mean and daily CV between 2020 and 2022. This experiment is in \url{sim_trace_analysis/change_over_time} directory, and the \url{calculate_mean_and_cv.py} script is run before plotting the results using \url{plot_change_over_time.py} script.
~\\\\
\textbf{Experiment (E3): [Periodicity Score]:} This experiment generates Figure 4, using the series period detect function from the Azure data explorer\footnote{\url{https://azure.microsoft.com/en-us/products/data-explorer}} to calculate the periodicity score of the carbon intensity signal. This experiment is in \url{sim_trace_analysis/periodicity} directory, and the \url{plot_periodicity_score.py} script is used to plot the periodicity scores. The sample periodicity score file is provided in  \url{sim_trace_analysis/periodicity/data_output}.
~\\\\
\textbf{Experiment (E4): [Region Capacity]:} This experiment generates Figure 5(a)-(c). The experiment script is \url{calculate_capacity.py}. Figures 5(a)-(b)  are plotted using \url{sim_spatial/geo_grouping_capacity/plot_geo_grouping_capacity.py} script, while Figure 5(c) is plotted using \url{sim_spatial/global_idle_capacity/plot_global_idle_capacity.py}
~\\\\
\textbf{Experiment (E5): [Capacity and Latency]:} This experiment generates Figure 6(a), where it sets the uniform idle capacity for all the regions in the dataset and limits the allowed latency that the workload can incur. We use the Google Cloud Inter-Region latency data for this experiment. This experiment is in \url{sim_spatial/capacity_latency} directory, and the \url{calculate_capacity_latency.py} script is run before plotting the results using \url{plot_capacity_latency.py}.
~\\\\
\textbf{Experiment (E6): [One and Infinite Migration]:} This experiment generates Figure 6(b), comparing the carbon reductions from the two migration policies, namely one-migration and infinite-migrations. For one migration, the workload is migrated to the region with the lowest annual mean and the job is run until completion. For the infinite-migrations policy, the workload is migrated to a region with the lowest carbon intensity of that hour, and the migration continues every hour until completion. This experiment is in \url{sim_spatial/one_and_inf} directory, and the \url{calculate_one_and_inf.py} script is run before plotting the results using \url{plot_one_and_inf.py} script.
~\\\\
\noindent
\textbf{Experiment (E7): [Vary Job and Slack]:} This experiment computes carbon emissions for deferrable and interruptible workloads for all 8760 potential start times over one year. The script to run this experiment is \url{sim_temporal/experiment_vary_job_slack.py}. The \url{sim_temporal/process_vary_job_slack_data.py} script then processes the raw output from the experiment. Both raw and processed results are stored in \url{sim_temporal/data_output} which will then be processed and plot in the other sub-directories. The processed data is then used to analyze carbon reductions from different dimensions in different sub-directories in the main \url{sim_temporal} directory. Note that the Weighted Workload load experiment involves getting the weights for each workload from the Google~\cite{google-trace} and Azure~\cite{azure} workload traces. To run the simulations in the \url{sim_temporal} sub-directories, go to \url{sim_temporal/<sub-directory>}, run the script that starts with \textit{calculate} to process the data, then run the script that starts with \textit{plot} to plot the results. This main temporal experiment together with the simulations in the sub-directories generate Figures 7(a)-(b), 8(a)-(b), 9(a)-(b), and 10(a)-(d).
~\\\\
\textbf{Experiment (E8): [Mixed Workloads]:} This experiment generates Figure 11(a), which analyzes how the proportion of the migratable workloads impacts carbon reductions. This experiment is in \url{sim_what_ifs/mixed_workloads} directory, run the \url{calculate_mixed_workloads.py}, then run \url{process_mixed_workloads.py} before plotting the results using \url{plot_mixed_workload.py}.
~\\\\
\textbf{Experiment (E9): [Prediction Error]:} This experiment generates Figure 11(b) that shows the increase in carbon emissions when there is a prediction error. To run the experiment:
\begin{itemize}
    \item Go to \url{sim_what_ifs/prediction_error} directory.
    \item Add prediction errors to the regions' error-free traces with \url{add_error.py} script.
    \item For each value of error, create a file that combines all the regions' traces with that error using \url{combined_spatial_trace.py}.
    \item Run the temporal shifting experiment using \url{temporal_shifting.py}
    \item Process the data from temporal shifting with \url{calculate_temporal.py}.
    \item Process the data for spatial shifting with \url{calculate_spatial.py}.
    \item Use \url{plot_global_scenario.py} script to plot the results.
\end{itemize}
~\\
\textbf{Experiment (E10): [Increasing Renewable Penetration]:} This experiment generates Figure 11(c)-(d) that shows the carbon emissions from carbo-agnostic and carbon-aware scheduling. To run the experiment:
\begin{itemize}
    \item Go to \url{sim_what_ifs/greener} directory.
    \item Create an emission factor file for all the available regions with \url{create_emission_factors.py}.
    \item Add more renewable sources to the raw carbon trace from Electricity Maps and re-calculate the carbon intensity with \url{add_renewables.py}.
    \item For each percentage of the added renewables, create a file that combines all the regions' traces with that percentage using \url{combined_spatial_trace.py}.
    \item Run the temporal shifting experiment using \url{temporal_shifting.py}.
    
    \item Process the results from temporal shifting using two scripts, namely \url{calculate_job_agnostic.py} and \url{calculate_job_aware.py}.
    \item Use \url{plot_temporal.py} and \url{plot_spatial.py} scripts to plot the results.
\end{itemize}
~\\
\textbf{Experiment (E11): [Temporal and Spatial Combined]:} This experiment generates Figure 11, which shows the combined reductions from temporal and spatial shifting. The sample regions are the same as the ones in \textit{(E3): [Periodicity Score]}. To run the experiment:
\begin{itemize}
    \item Access \url{sim_what_ifs/temporal_spatial_combined}.
    \item Calculate the carbon reductions from the \url{calculate_combined_savings.py} script.
    \item Use \url{plot_combined_savings.py} to plot the result.
\end{itemize}

\end{document}